\documentclass[apj]{emulateapj}

\usepackage{natbib}
\usepackage{mathtools}
\usepackage{graphicx}
\usepackage{multirow}
\usepackage{subfigure}
\usepackage{eucal}
\usepackage{url}

\slugcomment{Accepted for publication in ApJ 2015 July 29; published 2015 August 13.}

\begin{document}

\newcommand{\mt}{\ensuremath{\mathrm{M_\star }}}
\newcommand{\sfr}{\ensuremath{\mathrm{\dot{M}_\star }}}
\newcommand{\massunits}{\ensuremath{\mathrm{M_\odot }}}
\newcommand{\sfrunits}{\ensuremath{\mathrm{M_\odot}} \ensuremath{\mathrm{yr^{-1}}}}

\title{Unveiling the Milky Way: A New Technique for Determining the Optical\\ Color and Luminosity of our Galaxy}
\author{Timothy C. Licquia$^{1,2}$, Jeffrey A. Newman$^{1,2}$, Jarle Brinchmann$^3$}
\affil{$^1$
Department of Physics and Astronomy, University of Pittsburgh, 
3941 O'Hara Street, Pittsburgh, PA, 15260; tcl15@pitt.edu}
\affil{$^2$
PITTsburgh Particle physics, Astrophysics, and Cosmology Center (PITT PACC)}
\affil{$^3$
Leiden Observatory, Leiden University, PO Box 9513, 2300 RA Leiden, The Netherlands
}

\begin{abstract}

We demonstrate a new statistical method of determining the global photometric properties of the Milky Way (MW) to an unprecedented degree of accuracy, allowing our Galaxy to be compared directly to objects measured in extragalactic surveys.  Capitalizing on the high--quality imaging and spectroscopy dataset from the Sloan Digital Sky Survey (SDSS), we exploit the inherent dependence of galaxies' luminosities and colors on their total stellar mass, \mt, and star formation rate (SFR), \sfr, by selecting a sample of \textit{Milky Way analog galaxies} designed to reproduce the best Galactic \mt\ and \sfr\ measurements, including all measurement uncertainties.  Making the Copernican assumption that the MW is not extraordinary amongst galaxies of similar stellar mass and SFR, we then analyze the photometric properties of this matched sample, constraining the characteristics of our Galaxy without suffering interference from interstellar dust.  We explore a variety of potential systematic errors that could affect this method, and find that they are subdominant to random uncertainties.  We present both SDSS $ugriz$ absolute magnitudes and colors in both rest--frame $z$=0 and $z$=0.1 passbands for the MW, which are in agreement with previous estimates but can have up to $\sim$3$\times$ lower errors.  We find the MW to have absolute magnitude $^0\!M_r-5\log h=-21.00_{-0.37}^{+0.38}$ and integrated color $^0(g-r)=0.682_{-0.056}^{+0.066}$, indicating that it may belong to the green--valley region in color--magnitude space and ranking it amongst the brightest and reddest of spiral galaxies.  We also present new estimates of global stellar mass--to--light ratios for our Galaxy.  This work will help relate our in--depth understanding of the Galaxy to studies of more distant objects.

\end{abstract}


\section{Introduction} \label{sec:intro}
Galaxy evolution studies primarily rely on observational comparisons between objects in the local universe (e.g., the Milky Way; MW) and those at higher redshift, $z$ \citep[e.g.,][]{Faber07,Ilbert10,Leauthaud12}.  For most galaxies of known $z$, rest--frame colors and absolute magnitudes are some of the easiest global properties to measure, regardless of their distance from us.  Consequentially, color--magnitude diagrams (CMDs) provide a fundamental tool for interpreting galaxy evolution, especially at large $z$ where morphological information is difficult to obtain.  Yet to this day, the MW's position on such a diagram has remained poorly determined, despite being the galaxy we can study in the most detail.  Due to our location embedded within the disk of the Galaxy, interstellar dust obscures stars and hides most of the MW from our view \citep[cf.][]{Herschel}.  Furthermore, because bluer light is absorbed and scattered more efficiently by the dust, the optical colors of distant stars are altered \citep[e.g.,][]{Cardelli, SFD}.  Thus, determining the global optical properties of the MW from direct photometric observations has proven extremely difficult, requiring uncertain corrections and assumptions that are vulnerable to error.

For this reason, the history of measurements of the Galaxy's global photometric properties is sparse.  \citet[][hereafter vdK86]{vanderKruit} made the most recent significant measurement, utilizing a novel technique.  The Zodiacal cloud, a thick disk--shaped concentration of dust lying in the ecliptic plane (or zodiac), produces a glow of diffuse optical light throughout the night sky via the reflection of sunlight \citep{Reach}.  This diffuse glow, known as the Zodiacal light, introduces a significant amount of contamination to attempts to estimate the amount of starlight from the Galaxy.  The \textit{Pioneer} 10 spacecraft, launched in 1972 on a mission to Jupiter, became the first space probe to travel beyond the asteroid belt, to distances where the effect of the Zodiacal light becomes negligible.  van der Kruit used photometric measurements of the Galactic background light in broad optical blue (3950--4850\AA) and red (5900--6900\AA) bands taken by \textit{Pioneer} 10, corrected for diffuse Galactic light and extinction, and compared to stellar distribution models in order to find $M_B=-20.3\pm0.2$ and $B-V=0.83\pm0.15$ in the Johnson magnitude system.

Two earlier studies used a model of the Galaxy that consisted of a disk and spheroid component, but utilizing different data and assumptions, in order to infer the luminosity and color of the Galaxy; both of these yielded significantly bluer color estimates for the MW.  First, \citet[][hereafter dV\&P]{deVPence} had used a two--component model constrained to match the observed distribution of globular clusters in the Galactic bulge and the star counts near the Galactic poles of the disk in the solar neighborhood in order to infer $B-V=0.53\pm0.05$.  This work also yielded $M_B=-20.08$ (varying the shape of the bulge within this model yielded values ranging from -20.04 to -20.12); however, this estimate was updated to $M_B=-20.2\pm0.15$ in \citet[][hereafter dV83]{deV1983} assuming the Galactocentric radius of the Sun to be $R_0=8.5\pm0.5$ kpc.  Second, \citet[][herafter B\&S]{BahcallSoneira} constructed a similar model that combined a disk and spheroid component in order to match observed star counts as a function of magnitude, latitude, and longitude (rather than only the distribution of light across the sky).  This work yielded global values of $M_B=-20.1$, $M_V=-20.5$ and $B-V=0.45$ (assuming no reddening due to dust obscuration; no error estimates were provided).  dV\&P also summarized a series of earlier determinations of the Galaxy's total absolute magnitude made before its morphology and gross stellar structure were well understood.  These yielded estimates in the range of $M_B\simeq-19.5$ to $-19.9$, or $M_V\simeq-20.2$ to $-20.5$ \citep{Kreiken,deV1970,SchmidtKaler}, corresponding to an integrated $B-V$ color somewhere between 0.3 and 1.0 mag.  Additionally, dV83 averaged the colors from a set of galaxies believed to have nearly the same morphological type as the MW (assumed to be Sb/c) in order to infer $B-V=0.53\pm0.04$.

In the last decade, there has been a growing movement to quantify how typical the MW is amongst galaxies of its type \citep[e.g.,][]{Flynn, Hammer, Yin}; most of this work has used the measurements of vdK86.  We focus specifically on the recent work done by \citet[][hereafter M11]{Mutch} which investigates whether the MW is located in the so--called ``green valley'' \citep[cf.][and references therein]{Mendez11,Jin}, i.e., the sparsely populated region between the bimodal distribution of red and blue galaxies in the CMD \citep{Strateva01,Blanton03}.  After converting van der Kruit's measurement of Johnson $B-V$ to Sloan Digital Sky Survey (SDSS) AB model $u-r$ and placing the MW on a color--stellar mass diagram, M11 found that no secure conclusions as to the Galaxy's color could be drawn.  

To help resolve this question, we present in this paper a new method of determining our Galaxy's global photometric properties with dramatically smaller uncertainties.  Our technique resembles the ``sosies'' method utilized by \citet{sosies1} and \citet{sosies2}.  The underlying idea behind that technique was that if two galaxies match well in several calibrated properties, it can be assumed that they share the same luminosity, and hence differences in their apparent brightness can be used to determine their relative distances.  Here, we also look for sosies (i.e., analogs) of the MW; however, our goal is different, and we take advantage of larger datasets and more sophisticated statistical treatments in order to take into account uncertainties properly.  We derive our results using a method similar to that producing the dV83 value; i.e., we average the observed properties of galaxies selected as MW analogs, though here we carefully account for the systematic biases that can affect such an approach.

Essentially, we make the Copernican assumption that the MW should not be extraordinary for a galaxy of its stellar mass, \mt, and star formation rate (SFR), \sfr.  As these two properties are very strongly correlated with galaxies' global photometric properties, we first obtain a sample of MW analog objects that collectively match the stellar mass and SFR of our own Galaxy (taking into account the relevant uncertainties).  The range of observed photometric properties of galaxies in this sample provides tight constraints on our Galaxy's color and absolute magnitude.  With these values determined we are then able to accurately determine the MW's position in color--magnitude space.

Throughout this paper, all SDSS $ugriz$ magnitudes are reported on the AB system, whereas all Johnson--Cousins $UBVRI$ magnitudes are reported on the Vega system.  We use a standard $\Lambda$CDM cosmology with $\Omega_M=0.3$ and $\Omega_\Lambda=0.7$.  All absolute magnitudes are derived using a Hubble constant of $H_0=100h$ km s$^{-1}$ Mpc$^{-1}$, therefore making them measurements of $M - 5\log h$.  However, in order to compare measurements of the Galactic SFR and stellar mass, which are measured on an absolute distance scale, to those for extragalactic objects measured on the cosmic distance scale directly, we assume a Hubble parameter of $h=0.7$, following \citet{Brinchmann2004}.  Consequentially, the $\log\mt$ and $\log\sfr$ values we use for external galaxies can be adjusted to different choices of $H_0$ by subtracting $2\log(h/0.7)$.  For clarity, in what follows we will explicitly display the $h$--dependence of all quantities we use, as well as explain how our results for MW properties change with respect to $h$.

We structure this paper as follows.  In \S\ref{sec:data} we describe our observational data; this includes discussion of our total stellar mass and SFR estimates for the MW in \S\ref{sec:mw_sfr_mstar}, as well as discussion of the sample of externally measured galaxies we employ in \S\ref{sec:ext_gals}.  In \S\ref{sec:analysisSDSS}, we describe the criteria used in order to select the subsamples of SDSS galaxies used in this study.  In particular, we describe the selection of a sample of MW analog galaxies in \S\ref{sec:MWAS}, which we use to produce tight constraints on the integrated optical--wavelength properties of the Galaxy in \S\ref{sec:results}.  In \S\ref{sec:systematics} we investigate the principal sources of systematic error that may arise from our analog--sample selection methods.  We present our final results in \S\ref{sec:results}, including tables of useful photometric properties for the MW.  Lastly, we summarize this work and discuss its implications in \S\ref{sec:summary}.

\section{Observational Data} \label{sec:data}
In this section, we present a summary of the observational data we use for this study.  We begin by focusing on the total stellar mass, \mt, and SFR, \sfr, of the MW.  With these parameters in hand, we then describe the uniform parent sample of galaxies used, including the methods used to measure their stellar masses, SFRs, and rest--frame magnitudes.  The overarching goal of this study is to use this uniformly measured set of galaxies to convert our knowledge of the stellar mass and SFR of the MW into constraints on its global photometric properties.  The following section will detail how we construct a set of MW analog galaxies for that purpose.

\subsection{The Milky Way} \label{sec:mw_sfr_mstar}
In \citet[][hereafter LN15]{Licquia}, we present updated constraints on the total stellar mass and SFR of the MW, incorporating the wide variety of measurements in the literature.  For many of the same reasons that measuring the photometric properties of the MW is difficult (cf. \S\ref{sec:intro}), there are a limited number of estimates of both Galactic parameters in the literature.  In order to extract as much information as we can from these measurements, which encompass a variety of different observational data and methods, we employed a hierarchical Bayesian (HB) analysis method to combine all the measurements of a quantity into one aggregate result.  The HB method allows us to account for the possibility that any one of the included MW measurements is incorrect or has inaccurately estimated errors (e.g., due to neglected systematics).  The probability of erroneous measurements being incorporated into our meta--analysis is quantified by the inclusion of hyper--parameters in the Bayesian likelihood that characterize the data itself, and which we can simultaneously fit for along with the physical parameters of interest (e.g., \mt\ or \sfr).  The results of this study show that the conclusions from an HB analysis are robust to many different ways of modeling erroneous measurements.

With the present work in mind, the final results from LN15 are normalized so that they can be directly compared to the stellar masses and SFRs of external galaxies in the MPA--JHU catalog.  For the SFR of the MW, we capitalize on the work of \cite{Chomiuk}, which tabulated \sfr\ measurements made in the last three decades, renormalizing each to a uniform choice of the Kroupa broken--power--law initial mass function (IMF; \citealp{Kroupa}) as well as stellar population synthesis (SPS) code.  Applying the HB analysis method to these updated measurements yields a global SFR for the MW of $\sfr=1.65\pm0.19$ \sfrunits.  

For the total stellar mass of the MW, LN15 apply the HB analysis method to nearly 20 independent measurements of the stellar mass of the bulge component (including the contribution from the bar) from the literature, including results from photometric, kinematic, and microlensing techniques.  For the disk component of the Galaxy, we assume the single--exponential model from \cite{Bovy2013}; this is developed from the dynamical analysis of $\sim$16,000 G--type dwarf stars segregated into 43 mono--abundance populations based on their position in [$\alpha$/Fe]--[Fe/H] space, as measured by the SDSS/SEGUE spectroscopic survey.  Through Monte Carlo techniques we are able to simultaneously produce model--consistent realizations of the bulge and disk masses; we sum these two components to yield the total stellar mass of the Galaxy, \mt\ (the contribution of the stellar halo is negligible).  The Monte Carlo techniques allow us both to ensure that each bulge mass estimate is placed on equal footing and to incorporate the current uncertainties in the Galactocentric radius of the Sun, $R_0$.  In particular, we assume the constraints of $R_0=8.33\pm0.35$ from \citet{Gillessen}.  We show that once the bulge mass estimates are renormalized to the same definition of stellar mass (including main--sequence stars and compact remnants, but not brown dwarfs), scaled to the same $R_0$ appropriate to the measurement technique, and normalized to reflect consistent assumptions about the structure and demographics of the stellar populations (Kroupa IMF and single--exponential profile disk) then the results from our HB analysis are insensitive to models of potential systematics affecting the data.  All of this work culminates in a total stellar mass for the MW of $\mt=6.08\pm1.14\times10^{10}$ \massunits.

\subsection{SDSS Galaxies} \label{sec:ext_gals}
\subsubsection{Photometry} \label{sec:sdss_photometry}
To select a comparison sample of externally measured galaxies, we make use of data from the Eighth Data Release \cite[DR8;][]{DR8} of the Sloan Digital Sky Survey III \citep[SDSS--III;][]{York2000}.  DR8 provides both imaging and spectroscopic data for almost 10$^6$ galaxies in the local universe, spanning over a third of the night sky.  Its five broad optical passbands, labeled $u$, $g$, $r$, $i$, and $z$ in order of increasing effective wavelength, fully encompass of the CCD--wavelength window.  We make use of DR8, made available in early 2011, due to its best--to--date calibration and reduction of the imaging data.  All subsequent data releases from SDSS--III have provided no further refinements for low--$z$ galaxies as studied here.  The \textit{Photo} pipeline processing yields a variety of magnitude measurements based on fitting both a pure de Vaucouleurs and a pure exponential profile to the surface brightness distribution of each object.  Those quantities labeled ``\texttt{model}'' reflect the magnitude derived from the better of the two model profile fits in the best--measured band (generally $r$), which is then convolved with the object's point spread function (PSF) in each passband to obtain a template for measuring its flux.  DR8 also provides the magnitude derived from the optimal linear combination of the two model profiles that best fit the 2D image of any object in each passband, again convolved with the object's PSF; these are labeled ``composite model magnitudes'' or \texttt{cmodel}\footnote{See \url{http://www.sdss3.org/dr8/algorithms/magnitudes.php} for further detail, as well as discussions in \citet{Dawson}.}.  The \texttt{model} magnitudes are designed to produce the best, unbiased estimate of galaxy colors and so we use these to evaluate any color properties we discuss below.  However, while the \texttt{cmodel} magnitudes are not recommended for producing galaxy colors, they do reflect the best estimate of the ``total'' flux of a galaxy in each passband.  Therefore, all absolute magnitudes described below are derived from the \texttt{cmodel} measurements.

We have obtained $K$--corrections on all magnitudes in the DR8 catalog to rest--frame $z$=0 and $z$=0.1 SDSS passbands using the \texttt{kcorrect v4\_2} software package \citep{kcorrect}.  This entails fitting spectral energy distribution (SED) models to the observed $ugriz$ extinction-- and AB--corrected magnitudes, given the observed redshift, and then using this fit to determine offsets between observed quantities and magnitudes measured in rest--frame bands \citep[e.g.,][]{HoggKcorrection}.  The observed $z$ also provides a luminosity distance (given the cosmology we assume) and hence the distance modulus, $m-M$; we use the \texttt{kcorrect v4\_2} software to obtain rest--frame absolute magnitudes that are derived by subtracting this distance modulus along with the $K$--correction from the extinction-- and AB--corrected apparent \texttt{cmodel} magnitude in each band.  We obtain galaxy colors by taking the difference of two rest--frame absolute magnitudes, but using \texttt{model} magnitudes in place of \texttt{cmodel} as described above.  We choose to adopt the notation from \citet{kcorrect} when presenting our results: we denote an absolute magnitude for passband $x$ as observed at redshift $z$ by $^z\!M_x$.

At this point, we also use the \texttt{kcorrect} package to convert each galaxy's set of SDSS $ugriz$ (AB) magnitudes to an equivalent set of Johnson--Cousins $UBVRI$ (Vega) magnitudes, as well as their respective $K$--corrections.  This allows us to calculate $UBVRI$ extinction-- and $K$--corrected (\texttt{cmodel}--based) absolute magnitudes and (\texttt{model}--based) colors, which we can then analyze in parallel to $ugriz$ measurements in order to yield our results transformed to the Johnson-Cousin system.  As we will see in \S\ref{sec:summary}, this will be useful for comparing our results to the literature, and should be more robust than using any transformation equations available that are averaged over all galaxy types.

\subsubsection{MPA--JHU Stellar Masses and SFRs} \label{sec:mpa-jhu}
For a large sample ($\sim$10$^6$) of galaxies with spectroscopic redshifts from SDSS below 0.7, the MPA--JHU galaxy property catalog provides estimates of total stellar masses and SFRs.  These are currently publicly available at \url{http://www.mpa-garching.mpg.de/SDSS/DR7/}, and are based on SDSS Data Release 7 photometry.  However, to ensure the greatest possible accuracy in our results, for this study we have produced an upgraded version of this catalog by recalculating each galaxy's \mt\ and \sfr\ using the same algorithms, but applied to the photometric measurements released in DR8.  Hence, our initial dataset consists of the subset of galaxies in the MPA--JHU catalog that also have photometric measurements reduced through the DR8 pipeline.  All results presented herein are based on our DR8--based \mt\ and \sfr\ measurements, which assume a Kroupa IMF.  In the following, we briefly summarize the Bayesian methodology used to produce them.

Total stellar masses are determined following the same philosophy as \citet{Kauffmann} and \citet{Gallazzi05}, but by fitting models of SPS to each galaxy's photometry instead of using any spectral features.  Here, we first construct a large grid of galaxy models from \citet[][BC03]{BC03}, encompassing a wide range of possible star formation histories.  Each model produces a synthetic spectrum which we convolve with $ugriz$ passbands to produce model photometry.  For each galaxy, we then determine the likelihood for each model by calculating the $\chi^2$ from differences between fluxes corresponding to the model photometry and the observed \texttt{model} magnitudes.  Adopting flat priors on all model parameters, we then calculate the posterior probability for each model given the observations.  This is most similar to the methods of \citet{Salim}, differing in that the latter generated sets of input parameters by randomly drawing them from their priors instead of employing a grid.  Lastly, we integrate our grid of posteriors along all but the stellar mass axis in order to produce the marginalized posterior PDF for \mt.

SFRs are determined from the technique described in \citet{Brinchmann2004}, but with several improvements.  For star--forming galaxies this entails fitting the emission line models from \citet[][CL01]{CL01} to their H$\alpha$, \ion{O}{2}, H$\beta$, \ion{O}{3}, \ion{N}{2}, and \ion{S}{2} emission fluxes measured from their SDSS spectra, after subtracting the continuum and absorption features using the SPS spectra from the latest updates to the BC03 libraries.  In this case, a grid of $\sim$$2\times10^5$ CL01 models are investigated, which make up a four--dimensional grid of metallicities, ionization parameters, total dust attenuations, and dust--to--metal ratios.  Similarly as described above, the resulting grid of posteriors for all models can then be integrated over the other three axes to produce the marginalized posterior PDF for dust attenuation.  This is then used to estimate an unattenuated H$\alpha$ luminosity, which is then converted to a SFR using the \citet{Kennicutt98} conversion factor.

This yields a measurement of the SFR of each galaxy inside the SDSS 3$''$ fiber.  To overcome aperture bias, and hence produce an estimate of \sfr\ for the entire galaxy, we now follow in the footsteps of \citet{Salim}.  This requires calculating photometry for the light that falls outside of the fiber and fitting stochastic SPS models to it; for each galaxy we combine the SFR measured from inside and outside of the fiber to determine its total \sfr.  As a result, the SFRs employed herein should match well with the ``UV'' estimates by \citet{Salim} for all classes of galaxies over the entire dynamical range of \sfr\ values.  For a more in--depth discussion, the reader should see \citet{Brinchmann2004} and the MPA--JHU catalog website (listed above).

Ultimately, our methods produce DR8--based posterior PDFs for the log stellar mass and log SFR of each galaxy in our SDSS sample.  In our discussions to follow, when referring to a galaxy's $\log\mt$ or $\log\sfr$ we are truly referring to the mean value measured from the posterior.  We could equally have used median values for this study, as using them instead yields no differences in our results.  We also calculate the cumulative distribution functions (CDFs) measured from each galaxy's posteriors, and we use $\mathcal{P}_{x}$ to denote the value corresponding to the $x$th percentile in the CDF.  We then calculate an effective standard deviation for both quantities as $(\mathcal{P}_{84} - \mathcal{P}_{16})/2$.  We label these as $\sigma_{\log\mt}$ and $\sigma_{\log\sfr}$ hereafter.  We note that these effective error estimates are used only to screen galaxies with highly uncertain measurements in \S\ref{sec:init_cuts} and for investigating the impact of Eddington bias on our results in \S\ref{sec:EddBias}, and hence are sufficient for our purposes.

\subsubsection{Initial Cuts} \label{sec:init_cuts}
From the sample of galaxies we have described so far, we next restrict to a subset of those that make up the SDSS main galaxy spectroscopic sample (whose overall selection is described in \citealp{Strauss}), which includes only objects with good--quality, clean measurements.  To do so, we take advantage of the \textit{Photo} pipeline processing flags and image bitmasks to eliminate problematic objects from the full DR8 sample.  We first restrict to objects that were targeted as main sample galaxies by enforcing that the \texttt{primTarget} flag is set to ``galaxy.''  We then reduce to galaxies with good--quality observations taken from the \textit{Legacy} target plates by requiring the SDSS plate information tags \texttt{survey}, \texttt{programName}, and \texttt{plateQuality} are set to ``sdss,'' ``legacy,'' and  ``good,'' respectively.  We ensure a good--quality detection by requiring that the \texttt{BINNED1}, \texttt{BINNED2}, or \texttt{BINNED4} flag is set for the $r$--band image.  We exclude objects with any of the following $r$--band image flags\footnote{See \url{http://www.sdss3.org/dr8/algorithms/photo\_flags.php} and sources therein for explanations of these flags.} set: \texttt{SATUR}, \texttt{BRIGHT}, \texttt{BLENDED}, \texttt{NODEBLEND}, \texttt{DEBLEND\_NOPEAK}, \texttt{DEBLENDED\_TOO\_MANY\_PEAKS}, \texttt{PEAKCENTER}, \texttt{NOTCHECKED}, \texttt{CR}, \texttt{NOPROFILE}, \texttt{MANYPETRO}, \texttt{NOPETRO}, \texttt{PSF\_FLUX\_INTERP}, \texttt{BAD\_COUNTS\_ERROR}, \texttt{INTERP\_CENTER}, \texttt{BAD\_MOVING\_FIT}, or \texttt{DEBLENDED\_AT\_EDGE}.  At this point, we also exclude any galaxy whose inverse variance ($={1/\sigma^2}$) in absolute $g$-- or $r$--band magnitude is calculated to be 4 mag or smaller after $K$--corrections, or that has $\sigma_{\log\sfr} > 1$ or $\sigma_{\log\mt} > 0.5$, in order to exclude any object with highly uncertain luminosity, color, SFR, or stellar mass (these restrictions on property errors exclude only the most extreme outliers in the data, comprising $\ll1$\% of the sample).  As a result of these cuts, the DR8 sample is reduced to 337,331 galaxies from $\sim$10$^6$ with no restrictions applied.

\section{Constructing Useful SDSS Galaxy Samples} \label{sec:analysisSDSS}
We next trim our set of cleanly measured galaxies from the main galaxy sample to produce a uniform subset suitable for statistical analyses.  In this section we discuss the cuts employed to produce two important subsamples used in deriving our final results.  First, we describe the selection of a volume--limited sample, which consists of all galaxies lying in a redshift range such that any object with both SFR and total stellar mass values similar to those of the MW will be included in the SDSS sample.  Next, we discuss our method of identifying a set of MW analog galaxies from this volume--limited sample.  These galaxies, chosen based upon their \mt\ and \sfr\ values, can then be used to estimate the global photometric properties of the MW, while the volume--limited sample provides the context for discussing the MW's location in color--magnitude space.

\subsection{Selection of a Volume--limited Sample} \label{sec:VLS}
The SDSS main galaxy sample \citep{Strauss} is bounded by a limiting Petrosian magnitude of $r \le 17.77$ after correction for Galactic extinction.  Of course, this introduces a radial selection effect, known as Malmquist bias, whereby only the intrinsically brightest galaxies are present in the data at large $z$, whereas less luminous galaxies at the same redshift will not be targeted for spectroscopy.  We therefore restrict ourselves to a volume--limited sample, ensuring that all galaxies in the MW SFR and \mt\ ranges are detected and available for selection, regardless of their distances.

To do this, we first select a sample of MW analog galaxies via the process described in the following section from the clean main galaxy sample (i.e., ones with good--quality measurements, as described in \S\ref{sec:ext_gals}), but with no restriction in redshift.  Next, we overlay this sample of analogs upon the clean main galaxy sample with $z>z_{\text{min}}$ in $^0(g-r)$ vs. $^0\!M_r$ color--magnitude space; we then increase $z_{\text{min}}$ until, by eye, the included objects fall as faint as the faintest MW analogs, but no more so.  The corresponding value of $z_{\text{min}}$ will then serve as the \textit{maximum} redshift for our volume--limited sample.  This is true since we expect that the least--luminous galaxies in a sample must be at the smallest available redshift to be seen at all; or, we can conclude that at any redshift below $z_{\text{min}}$ such faint galaxies would be included in a sample, but at any greater redshift they would not.  Therefore, $z_{\text{min}}$ corresponds to the upper bound on the range of redshifts allowed for a volume--limited sample of MW analogs.

We note that this method is an extension of the standard procedure generally used to identify volume--limited samples of objects \citep[see, e.g.,][]{Tago10,Tempel14}.  Whereas one typically investigates the luminosity completeness level as a function of redshift, we have extended this analysis to the CMD.  In this way, we have ensured that all the results we present below (i.e., the luminosities and colors we infer for the MW) are guarded against Malmquist bias.  We also note that this process of choosing a $z_{\text{min}}$ contributes a negligible amount of uncertainty to our final results presented below.  For example, changing $z_{\text{min}}$ by $\pm$0.005 yields a shift in all of our results by $<0.05$$\sigma$.

Based on investigation of MW analog CMDs for different limiting redshifts, we find that a cut of $0.03 < z < 0.09$ ensures that all analogs have $r<17.77$, so that the SDSS magnitude limit has no effect on our results.  The lower bound on $z$ is used to limit the impact of aperture effects on the properties measured for these galaxies; again, a $\pm$0.005 shift in this value yields a $<0.05$$\sigma$ change in our results.  In addition to applying this redshift cut, we simultaneously enforce that all galaxy redshifts are measured at high confidence by ensuring that no redshift warning flags are set within the SDSS catalog (i.e., each has a value $\texttt{z\_warning}=0$).  The resulting volume--limited sample includes 124,232 galaxies from the clean main galaxy sample.
\begin{figure}[h]
\centering
\includegraphics[trim=0.35in 0.2in 0.65in 0.8in, clip=true, width=\columnwidth]{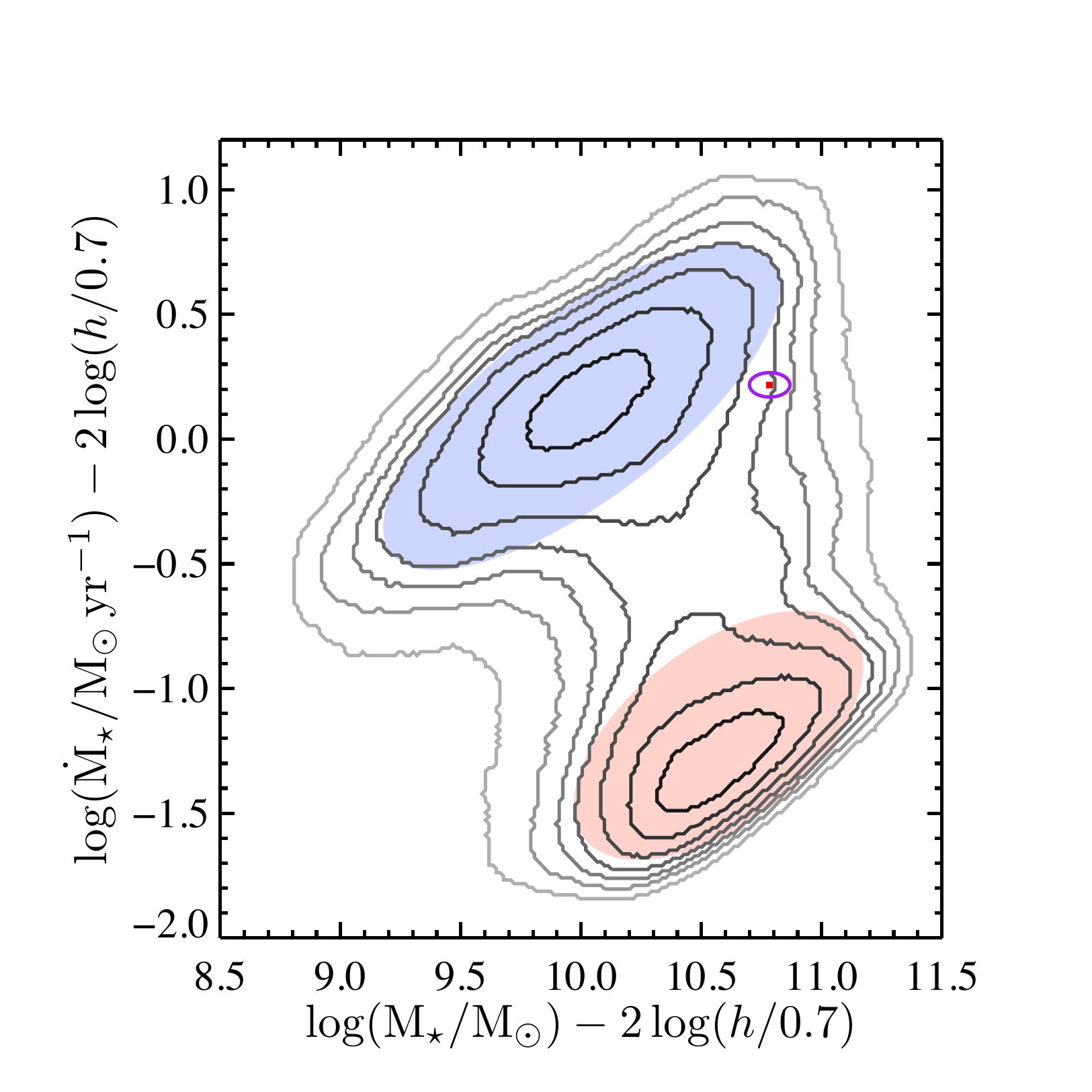}
\caption{Position of the Milky Way in star formation rate (\sfr) vs. total stellar mass (\mt) space.  Log--spaced contours depict the density of a volume--limited sample of SDSS galaxies in the range $0.03 < z < 0.09$.  The most likely position (red dot) and 1$\sigma$ constraints (purple) for the Milky Way shown here are determined from a hierarchical Bayesian meta--analysis of the literature estimates by LN15.  The properties for both the Milky Way and the extragalactic sample displayed here reflect consistent assumptions about their stellar populations, including a Kroupa IMF, and hence should be well guarded from any substantial systematics relative to one another.  We can see the Galaxy is offset from the main sequence of star--forming galaxies, hinting that it may be in a transitional evolutionary phase.}
\label{fig:MW_analytic_M*_SFR}
\end{figure}

In Figure \ref{fig:MW_analytic_M*_SFR}, we the show position of the MW with 1$\sigma$ constraints as determined by LN15 in \sfr--\mt\ space, overlaid upon log--spaced contours depicting the density of the volume--limited sample.  We have highlighted the approximate locations of the ``main sequence'' of star--forming SDSS galaxies in blue and the region of quiescent galaxies in red.  We also do this in our color--magnitude plots below; note that the relative positions of the regions corresponding to these two populations are flipped in magnitude space.  In the following section, we explain how we apply the Galactic constraints discussed in \S\ref{sec:mw_sfr_mstar} to the volume--limited sample in order to construct a sample of MW analogs, which in total should exhibit the same properties as the MW.  We can then examine where these analogs lie in color--magnitude space, ultimately converting our knowledge of where the MW lies in the \sfr--\mt\ plane into similar constraints on its photometric properties.

\subsection{Identifying Milky Way Analogs} \label{sec:MWAS}
We now collect a set of galaxies that, as an ensemble, can be used to constrain the overall photometric properties of our Galaxy; i.e., a sample of \emph{MW analog galaxies}.  By our definition, these analogs are selected in such a way that the distributions of their measured \mt\ and \sfr\ values match the posterior probability distributions describing the Galactic \mt\ and \sfr\ found in LN15 using a HB analysis (these results are detailed in \S\ref{sec:mw_sfr_mstar}).  To do so, we apply a randomized selection procedure to the galaxies in the volume--limited sample, as follows.

We begin by randomly drawing a single value from each of the adopted PDFs describing the MW's \mt\ and \sfr, independently of each other.  Ideally, we would like to then select a single galaxy from the volume--limited sample whose measured properties match these values exactly; we could then trivially build a sample of MW analogs by repeating this process a large number of times.  However, in general there will be no galaxies with properties that match these values perfectly.  Therefore, we use our pair of values drawn from the Galactic distributions as a point of reference in the \sfr--\mt\ plane and identify all galaxies from the volume--limited sample that lie within a small tolerance window centered on this point.  We choose this window to be the rectangular box that encompasses all values to within $x$\% of the \mt\ and \sfr\ values drawn.  To ensure that the distributions of analog properties will still match the fiducial Galactic posteriors, we require that $x$\% is much smaller than the error in either of the MW \mt\ and \sfr\ results presented in LN15.  Finally, from the galaxies that lie within our tolerance window we randomly select one as a MW analog.  We repeat this process 5000 times as the first step in building our sample, providing us with a set of 5000 MW analog galaxies.

We have employed a tolerance window in our method, as opposed to simply selecting the object in the volume--limited sample nearest each (\mt, \sfr) pair drawn, to maximize the number of unique MW analogs that make up our sample.  In practice, we find that when using a 1\% tolerance our window encompasses at least one galaxy from the volume--limited sample $\sim$75\% of the time, and typically contains up to eight candidate objects.  The remaining $\sim$25\% of the time we can expand our window to a 3\% tolerance from the drawn \mt\ and \sfr\ values in order to encompass a set of at least one galaxy, from which we randomly draw one analog.  Given that the fractional error in the adopted \mt\ and \sfr\ for the MW is $\sim$19\% and $\sim$12\%, respectively, we find that the exact size of the 1\%/3\% acceptance window is inconsequential to this study.  We have tested for the impact of using broader parameter space window sizes, though still small compared to the MW measurement errors, and always recover the same results (to well within the quoted errors).  We have also tested for any changes in our results when selecting analogs by their total specific SFR (sSFR) and \mt\ in place of SFR and \mt; again, the differences are well within the uncertainties.  In light of this, we have chosen to present the results of using the SFR measurements only, as using sSFR introduces substantial covariance (i.e., sSFR correlates strongly with \mt), whereas the SFR and total stellar mass of the MW are determined independently in LN15.

Just as it is a problem for observing the MW, dust alters the observed colors and magnitudes of star--forming galaxies observed with high inclination angles.  We therefore exclude objects likely to be edge--on spiral galaxies from our Milky Way analog sample (MWAS).  Accordingly, from the 5000 galaxies selected initially, we exclude all those that have both a best--fit axis ratio $b/a<0.6$ measured from a purely exponential profile fit to the surface brightness density in the $r$--band, as well as a value $f_{\text{deV}}<0.5$, where $f_{\text{deV}}$ effectively denotes the fraction of light in the galaxy's image that is contributed from a bulge--like component vs. a disk--like component, again generally measured from the 2D $r$--band image.  In effect, we are excluding all galaxies we have selected that are both edge--on and disk--dominated.  It is important to note that, because we have applied no morphological constraints on MW analogs, excluding disks in this way will introduce a morphological bias into our sample to some extent (i.e., the ratio of bulge--dominated and elliptical galaxies to disk--dominated galaxies will increase), an effect we will need to correct for.  Therefore, it is important to apply this cut \emph{only later} in our MW analog selection process so that we are able to track the fraction of disk galaxies that make it into the sample before and after its implementation; knowing these numbers will allow us to make the proper correction.  In \S\ref{sec:inclination} we will provide a more in--depth discussion of this, including how this inclination cut was chosen and its impact on our results.  Ultimately, after removing edge--on disks we are typically left with a clean sample of $\sim$3500 galaxies.
\begin{figure*}
\centering
\includegraphics[trim=0in 0in 0in 0in, clip=true, width=0.85\textwidth]{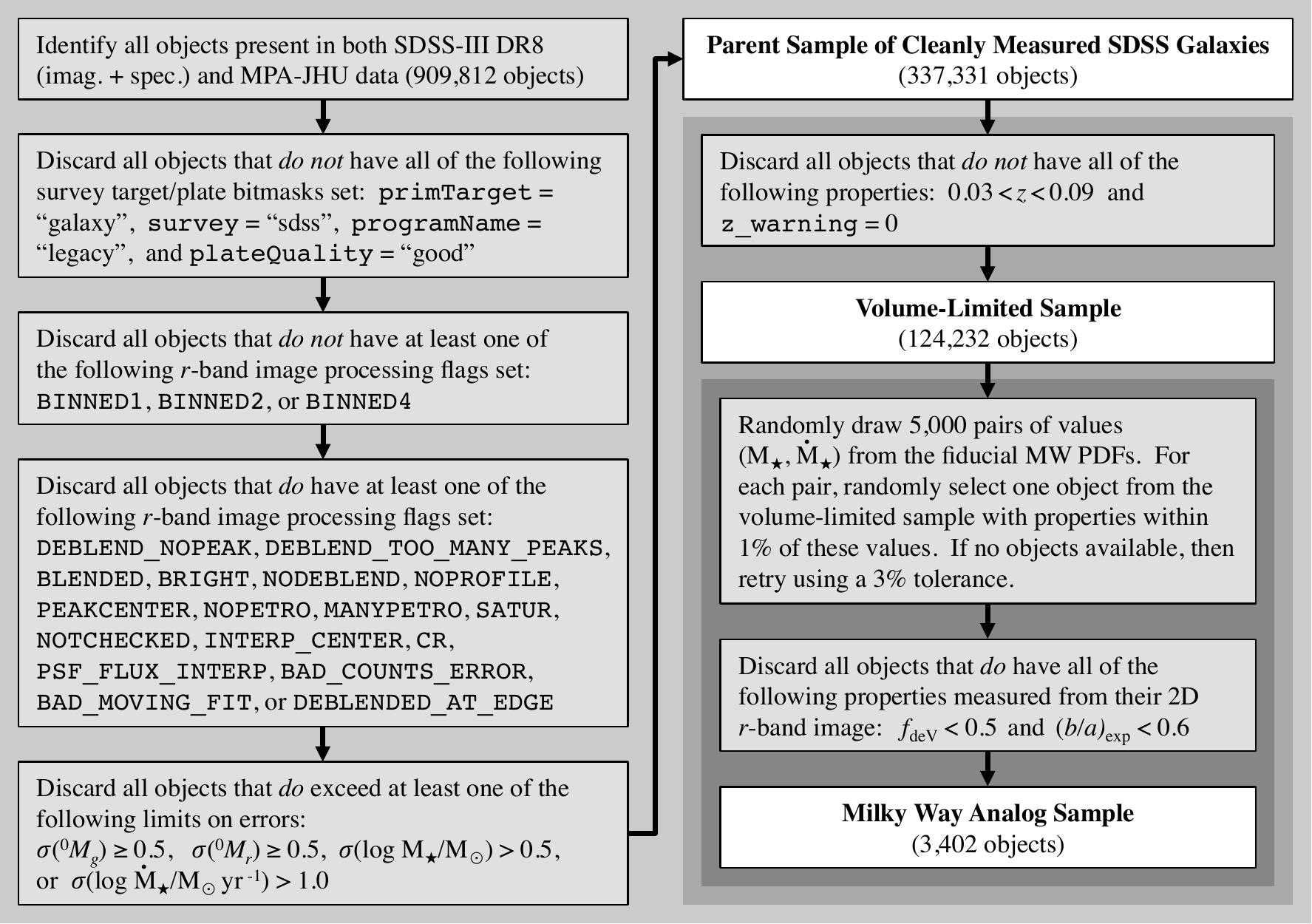}
\caption{Flowchart outlining the steps and criteria we use to select different samples of SDSS galaxies that we employ in this study.  This chart summarizes the processes described in sections \ref{sec:ext_gals}--\ref{sec:MWAS}, where more details may be found, including where we obtain or how we produce different property measurements for each object.  Note that here we denote the error in galaxies' stellar mass and SFR as $\sigma(\log\mt/\massunits)$ and $\sigma(\log\sfr/\sfrunits)$, respectively.  As mentioned in \S\ref{sec:MWAS}, $(b/a)_\textrm{exp}$ is the minor--to--major axis ratio obtained from the pure exponential profile best fit to an object's 2D image.}
\label{fig:flowchart}
\end{figure*}

For the particular realization we use in this study, our process more precisely yields 3402 galaxies that we will use to derive our results below, and which we henceforth call the MWAS.  We note that, of the analog galaxies selected, only 935 (or $\sim$27\%) are unique objects.  It is important to keep duplicate objects so that the distribution of property values for the MWAS accurately matches the posterior distributions for the MW properties we have found in LN15 (see \S\ref{sec:mw_sfr_mstar}).  In practice, we find that if we only keep the set of unique objects, the mean \mt\ of our sample has a significant bias ($\sim$2$\times$ the standard error) compared to when we eliminate duplicates.  The SFR distribution is affected less.  These biases are avoided altogether by allowing objects to be selected multiple times as a MW analog.

\begin{figure}[h]
\centering
\includegraphics[trim=0.35in 0.25in 0.65in 0.8in, clip=true, width=\columnwidth]{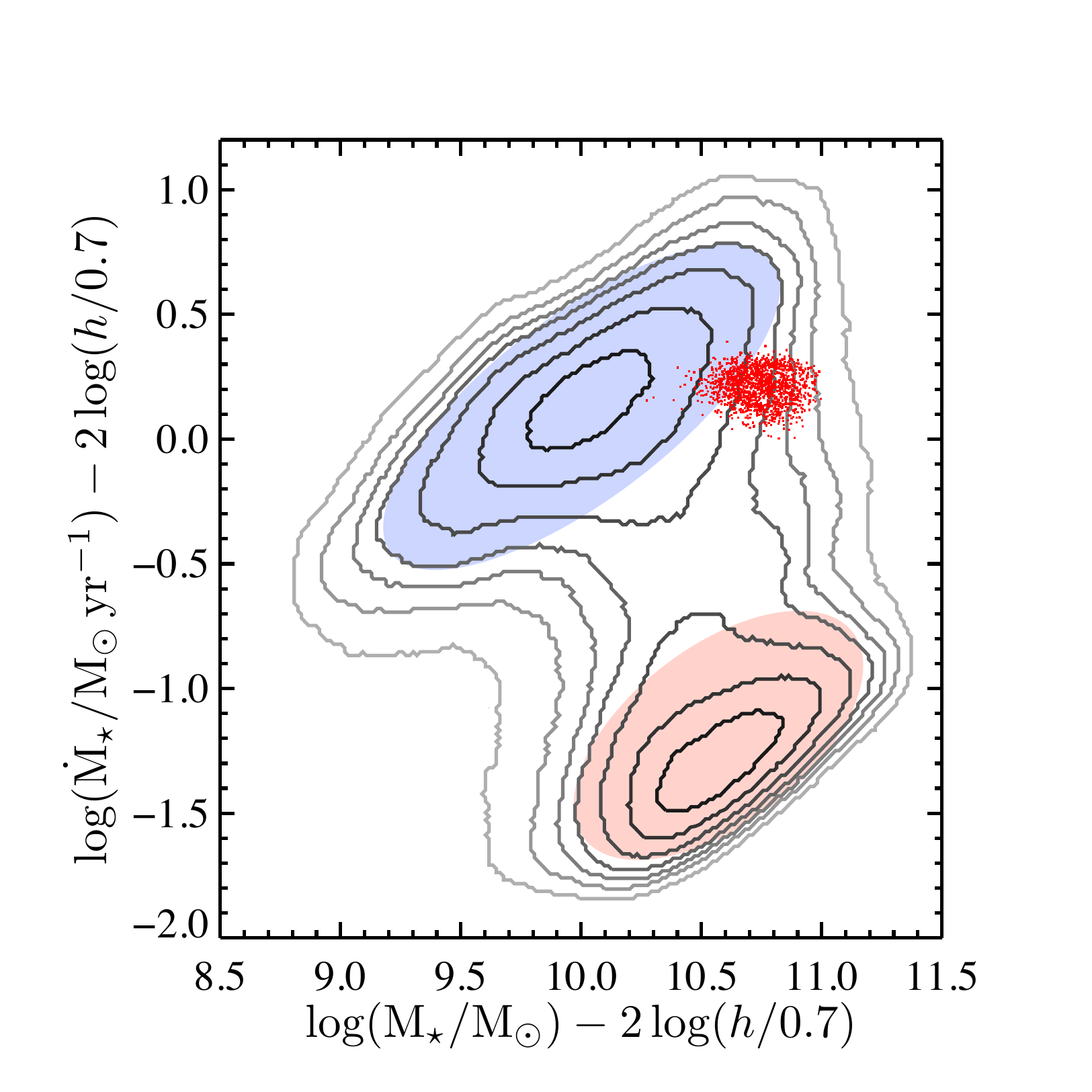}
\caption{Sample of $\sim$3500 Milky Way analog galaxies (red dots), chosen through a random selection process such that they collectively match the same distribution in SFR--\mt\ space as the Milky Way (compare to Figures \ref{fig:MW_analytic_M*_SFR} and \ref{fig:mstar_sfr_hists}).  The grayscale, log--spaced contour lines depict the density of a volume--limited sample of SDSS galaxies ($0.03 < z < 0.09$) that encompasses the Milky Way SFR and \mt\ ranges throughout this redshift range, but is not affected by any limiting magnitude.  This is the same sample from which the Milky Way analogs are drawn (see \S\ref{sec:ext_gals} for the details).  Fundamentally, we make the Copernican assumption that the Milky Way should not be extraordinary amongst the set of galaxies of similar stellar mass and star formation rate, and hence \emph{some} galaxy in that set must have closely matching photometric properties; in this study we focus on integrated optical--wavelength properties which are all but impossible to measure directly.}
\label{fig:MW_analogs_M*_SFR}
\end{figure}

For convenience, we show a flowchart in Figure \ref{fig:flowchart} that summarizes sections \ref{sec:ext_gals}--\ref{sec:MWAS} into a step--by--step procedure that yields all of the different samples of galaxies that we employ in this study.  Figure \ref{fig:MW_analogs_M*_SFR} shows the positions of the MWAS in \sfr--\mt\ space as red dots, overlaid upon the same contours for the volume--limited sample as Figure \ref{fig:MW_analytic_M*_SFR}.  The spread of these dots appears broader than the Galactic constraints in Figure \ref{fig:MW_analytic_M*_SFR} due to both the saturation of color where there are many objects and the substantial number of $>$3$\sigma$ events to be expected in any sample of 3500 numbers; as mentioned above, the size of the search box is small in comparison to the spread in MW values.

\begin{figure}[h]
\centering
\includegraphics[trim=0.5in 0.25in 0.8in 0.8in, clip=true, width=\columnwidth]{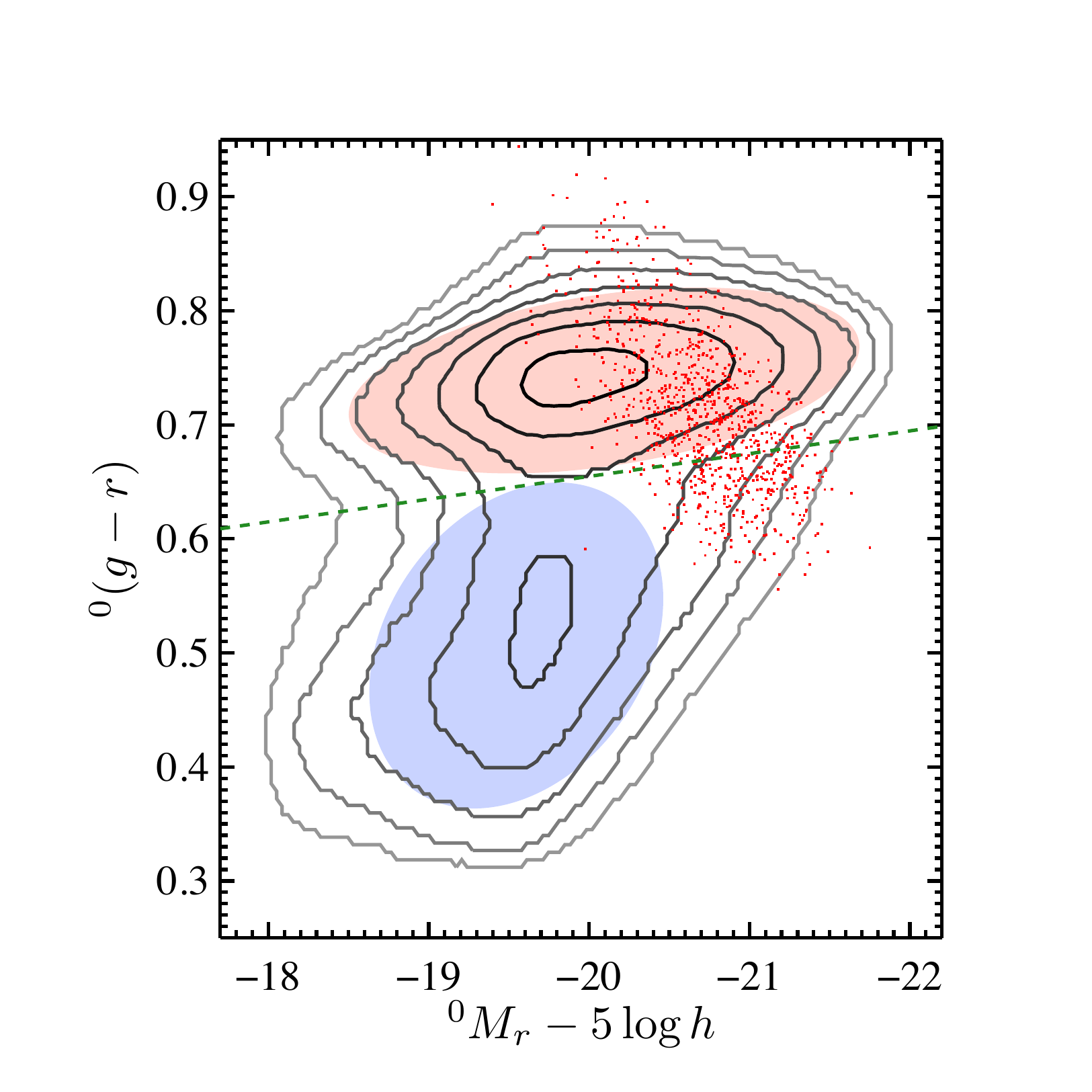}
\caption{Sample of $\sim$3500 Milky Way analog galaxies (red dots) plotted in SDSS $^0(g-r)$ vs. $^0\!M_r$ space.  This is the same sample of objects shown in Figure \ref{fig:MW_analogs_M*_SFR}; i.e., they are selected to produce a distribution of star formation rate and total stellar mass values matching the probability distribution describing those properties for our Galaxy.  Again, the grayscale, log--spaced contours depict the density of a volume--limited sample of SDSS galaxies in the range $0.03 < z < 0.09$ (see \S\ref{sec:VLS}).  For reference, the dashed green line shows a simple SDSS color cut dividing the red--sequence and blue--cloud regions (G. \citeauthor{Graves} 2012, private communication; see \S\ref{sec:MWAS} for more details).}
\label{fig:MW_analogs_cmd}
\end{figure}

Figure \ref{fig:MW_analogs_cmd} shows our sample of MW analogs (red dots) overlaid on the volume--limited sample (grayscale contours), similar to Figure \ref{fig:MW_analogs_M*_SFR}, but now plotted in the $^0(g-r)$ vs. $^0\!M_r$ CMD.  Mapping these galaxies from one parameter space to the other noticeably increases their scatter compared to the underlying population from which they were drawn.  However, their tight correlation in \sfr--\mt\ space, as expected, produces significant constraints in the CMD, providing us with information on what locations could feasibly be occupied by the MW.  We display a division line between the red sequence and blue cloud regions obtained by taking a line parallel to the slope of the red sequence, but offset to the point where contributions from red sequence and blue cloud galaxies are equal, determined based upon the assumption that the spread in red sequence colors about the center line is Gaussian (G. \citeauthor{Graves} 2012, private communication; cf. \citealp{Taylor14}).  The peak density of the MW analogs lies near the division line, though many lie far above or below it.

With the MWAS in hand, we are now ready to calculate new constraints on the photometric properties of the MW.  For instance, simply calculating the mean and standard deviation of the $^0(g-r)$ colors of our sample yields $\sim0.72\pm0.07$, and similarly we find $\sim-20.75\pm0.37$ for $^0\!M_r$.  However, as mentioned above, before presenting our final results we must first account for any major potential sources of systematic error in our method (e.g., the morphological bias introduced from removing any edge--on disks from the MWAS), and make the proper corrections.  We will next discuss these systematics and the corrections that they require in the following section.

We note that for any \textit{mean} quantity described hereafter, including those provided in our tables of results, we are actually using the Hodges--Lehmann (H--L) estimator \citep{HLMean}.  The H--L estimator is a robust measure of the median of the data, which is calculated by determining the median value of the set $\left\{(x_i+x_j)/2\right\}$ for all pairs $i,j$.  For $N$ Gaussian--distributed data points with a standard deviation $\sigma$, the mean has standard error $\sigma/\sqrt{N}$, while the median has uncertainty $\sim$$\sigma/\sqrt{0.64N}$.  The H--L estimator has an error of $\sim$$\sigma/\sqrt{0.955N}$, comparable to the mean, but shares the robustness to outliers of the median, making it a superior choice in most cases.  Hence, using the H--L estimator should reduce the impact of significant outliers in our sample, in contrast to the ordinary mean, but it will have smaller errors than the median.

Instead of calculating for all possible pairs, which requires excessive computation time, we bootstrap this estimate by choosing a set of random $i,j$ pairs equal to ten times our effective sample size (reducing from $\sim6\times10^6$ total pairs down to a much more manageable 34,020 for our typical calculations).  This introduces a small amount of extra uncertainty ($=\sigma/\sqrt{2\times0.64\times34,020}$) which must be added in quadrature to the nominal standard error in the H--L estimator.  The net result is that our estimator yields uncertainties 3\% larger than the true H--L mean would.  This additional uncertainty is negligible compared to our overall errors; hence this technique does not introduce any measurable amount of potential bias, and the bootstrapped H--L estimator in our application will still have significantly smaller uncertainty than the median value.

\section{Systematics} \label{sec:systematics}
In this section we discuss the principal systematic errors and biases that could affect the methods applied in this study, other than systematic errors in either of MW or extragalactic \mt\ and \sfr\ measurements, which we defer discussion of to \S\ref{sec:results}.  First, we investigate the impact of Eddington bias, i.e., the bias resulting from selecting objects using quantities that are affected by measurement errors.  We provide details on how we can estimate its overall effect, which we then subtract from our final results.  In addition, we analyze the impact of reddening associated with observing disk galaxies at an inclination on the optical properties of our MWAS.  We discuss how we identify inclined objects in the SDSS measurements, as well as our methodology for mitigating reddening or extinction effects that otherwise, when neglected amongst the MWAS, could distort the inferred photometric properties of our Galaxy.    We will demonstrate that, after corrections for these effects, the remaining systematic uncertainties from these effects are well below statistical uncertainties.
\begin{figure*}
\centering
\begin{subfigure}
\centering
\includegraphics[trim=0.35in 0.2in 0.8in 0.8in, clip=true, width=\columnwidth]{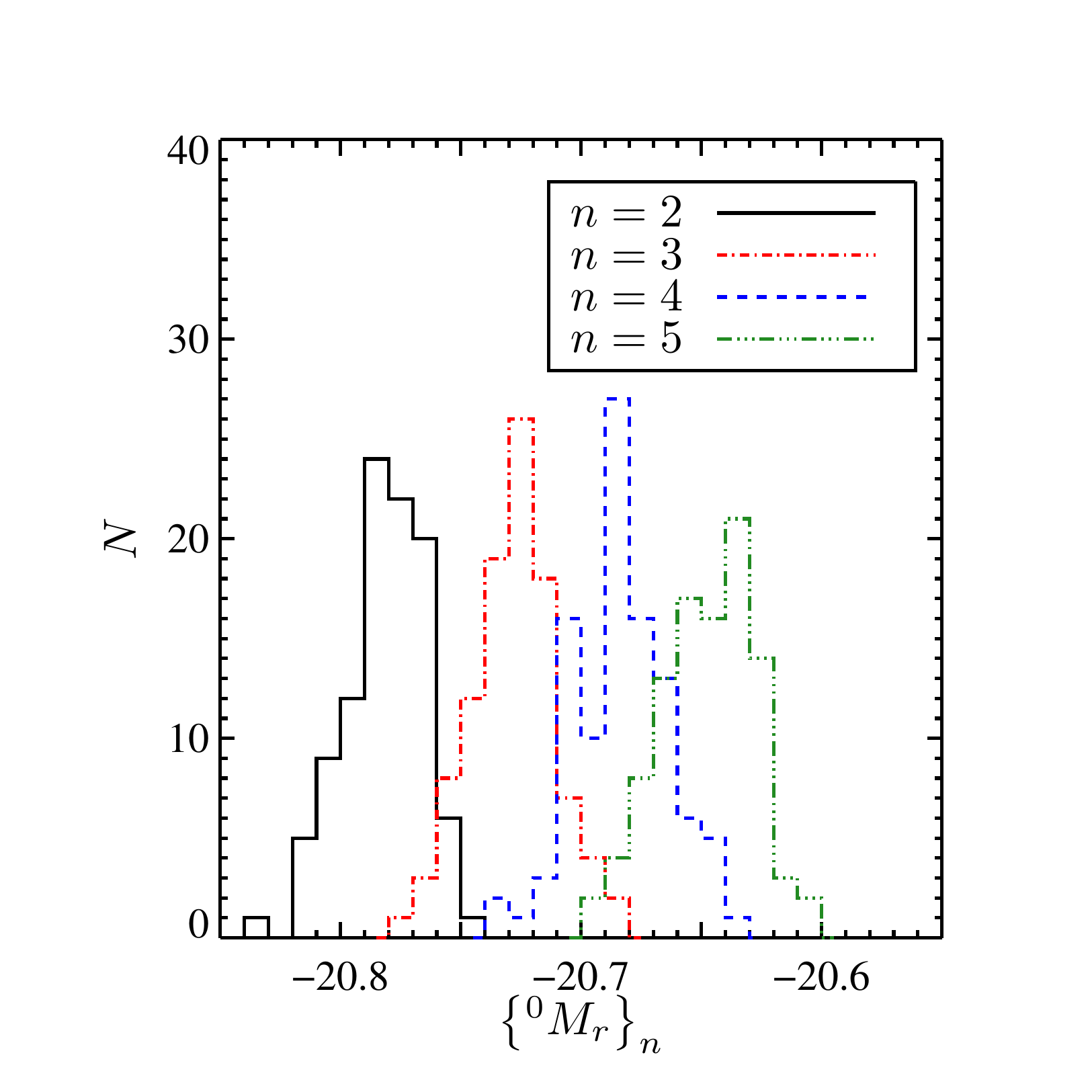}
\end{subfigure}
\begin{subfigure}
\centering
\includegraphics[trim=0.35in 0.2in 0.8in 0.8in, clip=true, width=\columnwidth]{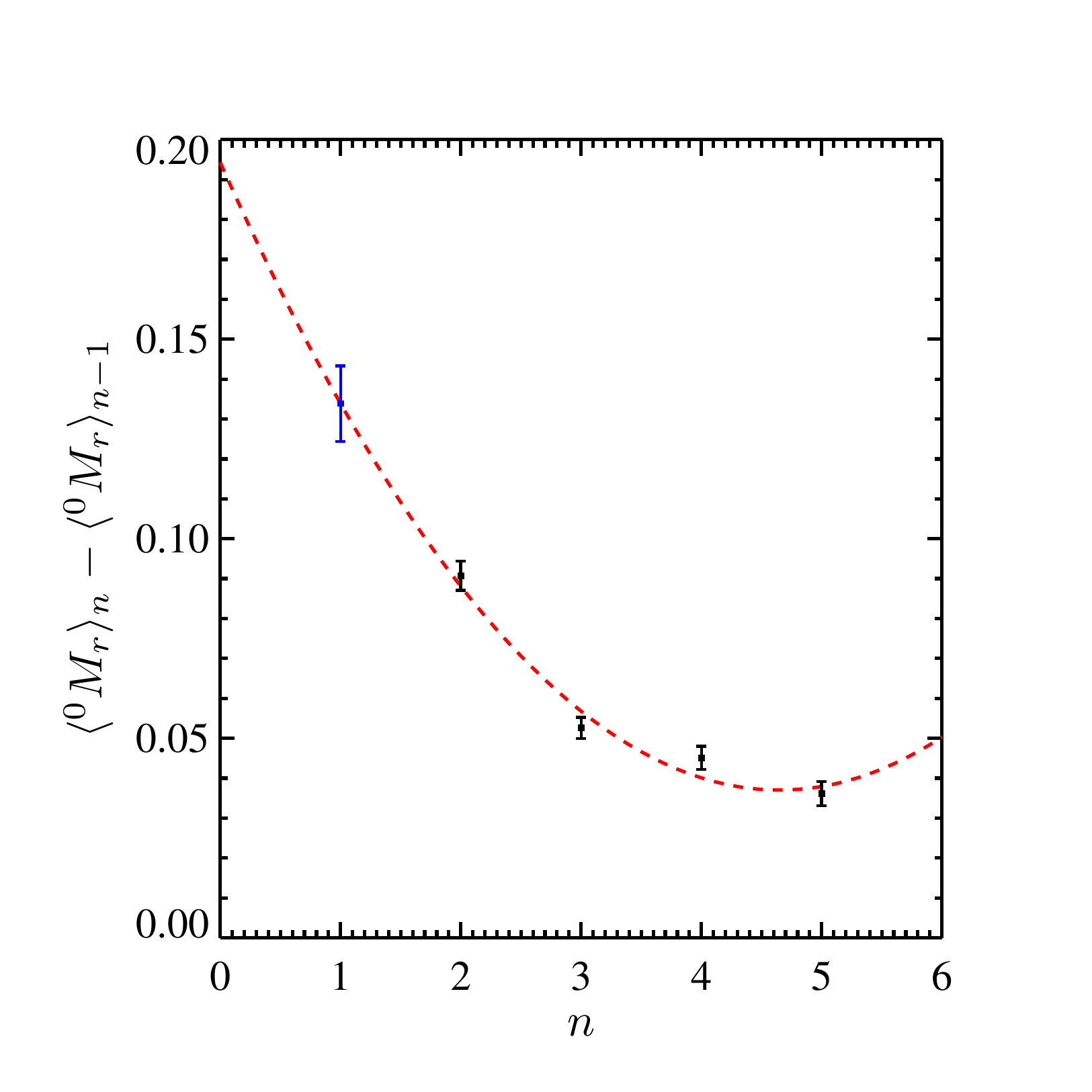}
\end{subfigure}
\caption{Modeling the Eddington bias in the Milky Way analog selection method.  \textit{Left panel:} histograms of the mean absolute $^0r$--band magnitudes produced from Monte Carlo (MC) simulations of selecting a new sample of Milky Way analog galaxies as increasing amounts of noise are added to galaxies' total stellar mass (\mt) and star formation rate (\sfr) values.  This noise is drawn from a normal distribution, with mean of zero and standard deviation determined by the errors in a galaxy's estimated \mt\ and \sfr, and is applied $p$ times successively before analogs are selected.  Since nominal values are all affected by noise, we denote them as \emph{the} $n=1$ case, and so any further degradation is marked $n=p+1$.  \textit{Right panel:} a least--squares quadratic fit to the four points yielded by subtracting the mean of $\left\{^0\!M_r\right\}_{n-1}$ from that of $\left\{^0\!M_r\right\}_n$, using the distributions from the left panel, as a function of $n$.  $\langle^0\!M_r\rangle_1$ is measured from the mean $^0\!M_r$ of our MWAS to produce the point at $n=2$.  We then use this fit to extrapolate the blue datapoint at $n=1$, the ordinate of which should reflect the difference between the actual measurements, which are affected by Eddington bias, and what would be measured with zero errors, i.e., the quantity we desire; this value is subtracted from the observed absolute $r$--band magnitude of the sample.  This same process is applied to each absolute magnitude or color considered in this study, and the bias subtracted is listed in Tables \ref{table:props1}--\ref{table:props4}.  Almost always this offset is completely subdominant to the statistical errors of our method; the exception is $u$--band--based color measurements, for which the bias is of the same order as, but still smaller than, statistical uncertainties.  Even then, the uncertainty in the bias correction is much smaller than other sources of error.}
\label{fig:edd_bias}
\end{figure*}

\subsection{Eddington Bias} \label{sec:EddBias}
It is important to address how the uncertainties in our stellar mass and SFR estimates affect our results.  Specifically, we are drawing each MW analog from a small bin in SFR--\mt\ space.  For any parameter whose intrinsic probability distribution function has significant higher derivatives (second or beyond), scatter due to errors will move more objects from bins with more objects to those with fewer as opposed to the converse.  This causes the observed distribution of values with errors to be biased compared to the true, underlying distribution.  This phenomenon is known as Eddington bias and is very common in astronomy; it is the generalized form of the Malmquist bias that affects luminosity distributions.  For instance, since massive galaxies are rare, a galaxy with a large \mt\ estimate is more likely to have an actual stellar mass below that value than above, since there are many more objects that could up--scatter than down--scatter.  As a result, in aggregate the \mt\ values of our MW analogs should be biased high.  Similar effects could affect SFR, luminosity, or color estimates.

To quantify this bias, we consider a statistical exercise of perturbing each galaxy's \mt\ and \sfr\ values by Gaussian noise sampled from their estimated errors, and then reselecting a set of MW analog galaxies utilizing the perturbed measurements.  To be specific, we offset the mean $\log\mt$ and $\log\sfr$ values individually for each galaxy by a value randomly drawn from a Gaussian distribution centered at zero with a standard deviation of that object's $\sigma_{\log\mt}$ or $\sigma_{\log\sfr}$ value.  We perform $N=100$ realizations of this perturbation process, each time selecting a new set of MW analog galaxies in the same manner described in \S\ref{sec:MWAS}.  Calculating the H--L mean for each $ugriz$ property for each realization yields a distribution characterizing our nominal results with the effects of Eddington bias applied twice, instead of the single impact that should affect our standard sample.  We bootstrap this distribution of doubly biased values to measure the mean H--L mean and its standard error \citep{efron}.  We then repeat this exercise but applying the noise 2, 3, or 4 times consecutively before selecting a sample.  This yields distributions of the mean property of interest after repeatedly applying the bias in our method $n$ times; the mean of this distribution we denote $\mu_n$.  For clarity, note that we consider the actual Eddington bias in our standard MWAS as the first ($n=1$) application, and thus distributions of $ugriz$ properties yielded from $p$ successive perturbations of the \mt\ and \sfr\ values by their errors are labeled $n=p+1$ in our plots and discussion below.  Figure \ref{fig:edd_bias} displays examples of this analysis for $^0\!M_r$.

To estimate the Eddington bias in each property we then plot the difference between the means of the $n$ and $n-1$ values of a given parameter as a function of $n$; i.e., $\mu_n-\mu_{n-1}$ versus $n$. We then perform a least--squares quadratic fit to these four data points, incorporating the error estimates from our bootstrap analysis.  We use the resulting curve to extrapolate to $n=1$, whose ordinate corresponds to the offset in the ``mean'' of a given property between when Eddington bias affects our sample of MW analog galaxies and when it does not.  This value is then subtracted from the observed mean for that property of the MWAS.  In Figure \ref{fig:edd_bias} we show what the results of this exercise typically look like, again adopting $^0\!M_r$ as an example.

In order to calculate the uncertainty in our estimate of the Eddington bias, we construct the covariance matrix for the coefficients of a least--squares quadratic fit, $A+Bn+Cn^2$.  We are interested in the $\sigma$ of the point at $n=1$; this simply reduces to the square root of the sum of all elements of the covariance matrix.  We note that if the uncertainties in our stellar mass estimates were primarily due to photometric errors, this treatment would be incorrect, because if an object had (say) a higher--than--actual estimated \mt\ value, it would also have a too--bright $M_r$.  However, this does not appear to be the case; we find that stellar mass errors are $>5\times$ larger than would be expected from SDSS photometric errors, so other sources of uncertainty clearly dominate, and we can safely treat absolute magnitudes and stellar masses as statistically independent.

\subsection{Inclination Reddening} \label{sec:inclination}
As mentioned in \S\ref{sec:MWAS}, we have removed any edge--on disk galaxies that originally were included in the MWAS.  This is because, just as reddening and extinction affect observations though the disk of the MW, they also alter measurements of external spiral galaxies with their disks aligned along our line of sight.  The SED measured for such an edge--on disk galaxy will be significantly distorted; we will receive a much smaller fraction of blue light than when observing face--on, while the flux of redder optical light detected will be much less affected.  This means that the more edge--on a disk galaxy in the SDSS sample is, the less representative our observations will be of its intrinsic photometric properties \citep[see, e.g.,][]{Inclination,Maller,Salim09}.  It is important, therefore, to ensure the properties of the MWAS are not skewed by this effect.  In this section we discuss quantitatively our method of choosing the appropriate inclination threshold for the MWAS, as well as the unwanted side effect that it creates, namely morphological bias, which we will also need to correct for.

DR8 provides measurements of each galaxy's ratio of semiminor to semimajor axis, $b/a$, as determined from the exponential profile best fit to its 2D image (labeled \texttt{abExp} in the DR8 catalog).  Low values of $b/a$ indicate that the galaxy's image has high eccentricity.  Additionally, when calculating \texttt{cmodel} magnitudes (see \S\ref{sec:ext_gals}), the weight of the de Vaucouleurs profile in the best--fit linear combination with the exponential profile matched to the object's image is recorded as $f_{\text{deV}}$ (or alternatively \texttt{fracDeV}).  Essentially, this quantifies the fraction of the total light in the 2D image of the galaxy that is well fit by a S\'ersic index of 4 as opposed to 1.  For our purposes, we classify any object with $f_{\text{deV}}>0.5$ as a bulge--dominated or elliptical galaxy, and any with $f_{\text{deV}}<0.5$ as a disk--dominated galaxy.  Our objective is then to use these two parameters to effectively eliminate edge--on disk--dominated objects (i.e., ones well fit by an exponential profile and that appear to have a high inclination angle) from being selected as part of the MWAS to prevent any systematic offsets in our results due to reddening.  We are not concerned about including early--type galaxies with low $b/a$ in the sample, as they contain little cold gas and dust, so extinction effects are comparatively minor for them.

\begin{figure}[t]
\centering
\includegraphics[trim=0.4in 0.25in 0.7in 0.8in, clip=true, width=\columnwidth]{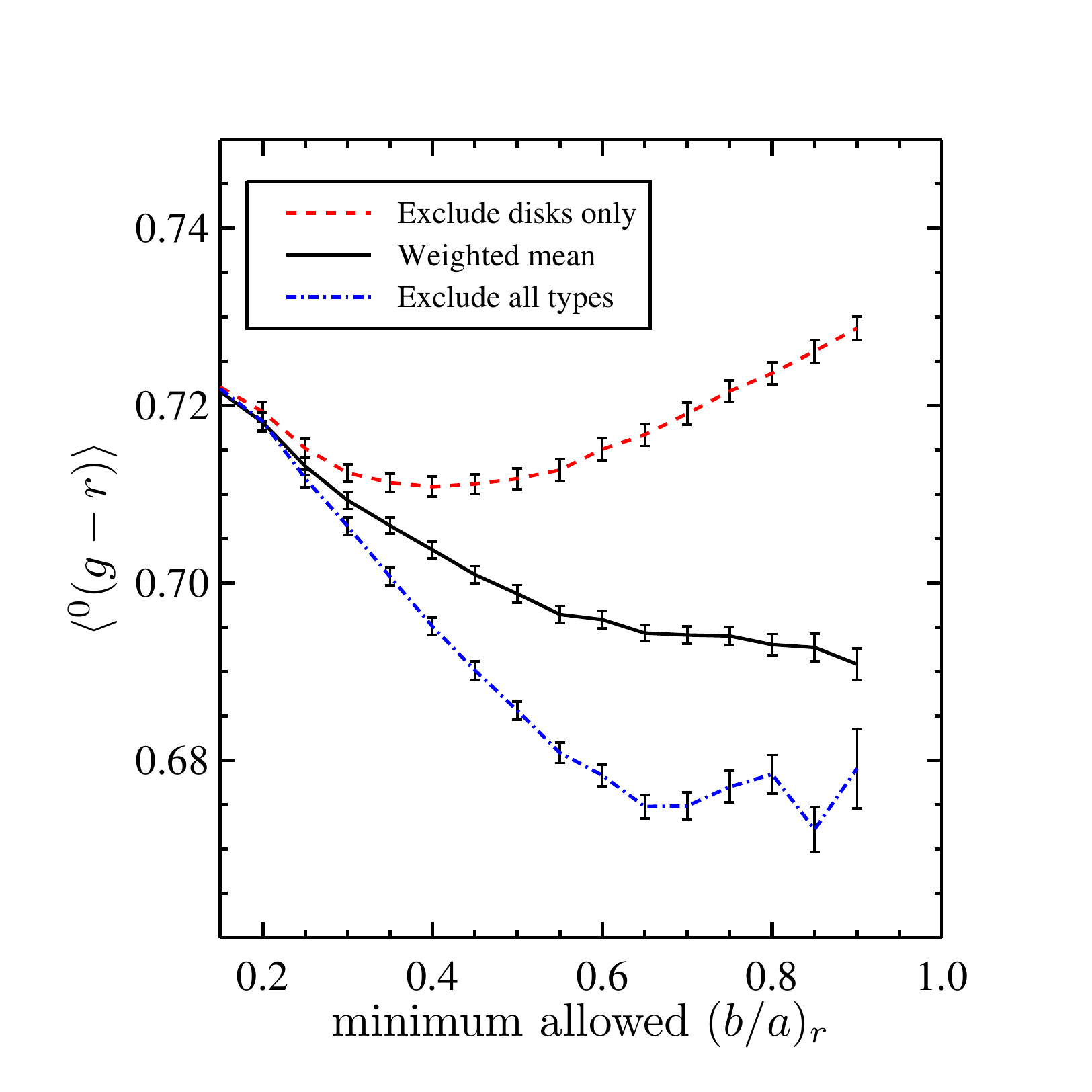}
\caption{Mean $^0(g-r)$ color of our Milky Way analog sample as a function of the minimum allowed axis ratio of the objects included in three different scenarios.  The red dashed curve is the result of removing only disk galaxies (i.e., objects having $f_{\text{deV}}<0.5$) based on the axis ratio cut given on the $x$--axis.  The blue dashed--dotted curve is the result of removing any galaxy, regardless of type, based on the minimum ratio cut.  The solid black line reflects the result of removing only disk galaxies again, but in this scenario we add extra weight to the contribution from disk galaxies remaining after the cut to the overall mean $^0(g-r)$ color to correct for the objects removed.  Excluding only disk galaxies initially causes a trend toward net bluer color as edge--on systems are removed, but eventually this trend reverses toward redder average color due to increasingly oversampling the elliptical population.  However, in the second case, since spheroids outnumber disks at $b/a\gtrsim0.35$, throwing away any galaxy above this minimum allowed threshold means discarding more spheroids than disks, giving extra weight to the blue population.  Therefore, we chose the last scenario (a weighted mean) as our fiducial method, as it provides a stable behavior over a large range of reasonable choices of cutoff for $b/a$, and hence appears robust to such morphological bias.  That is, the slope of the black curve is shallow enough that moving from a minimum allowed $b/a$ of 0.4--0.8 would cause a $\lesssim0.01$ magnitude change; $b/a>0.6$ is our fiducial cut.  The other scenarios provide a much more crude and extreme way of removing inclination reddening from our sample; we note that the offset of the other curves from the black curve at $x=0.6$ is still $\lesssim0.02$ mag, which is subdominant to the statistical error ($\sim$0.06 mag).  The analogous plot for $^0r$--band absolute magnitude yields a similar conclusion, and so we adopt the weighted--mean scenario as standard for all quantities.}
\label{fig:axis_ratio}
\end{figure}

We have tested the impact of this cut by varying the minimum allowed axis ratio for disk--dominated galaxies in our sample and measuring how the mean $^0(g-r)$ color is affected.  To do so, we again employ a process of selecting a set of 5000 MW analog galaxies identical to that explained in \S\ref{sec:MWAS}, but unlike before, we do not yet apply any constraints on the $b/a$ or $f_{\text{deV}}$ values in the sample.  In Figure \ref{fig:axis_ratio}, we display how the integrated color of this set changes as we impose constraints on $b/a$ and $f_{\text{deV}}$ in three different ways.

First, the upper (red dashed) curve shows the effect of removing only disk galaxies (i.e., ones having $f_{\text{deV}}<0.5$) as we systematically increase the minimum allowed axis ratio.  Initially, as we increasingly remove the lowest $b/a$ (most inclined) disks we see the mean $^0(g-r)$ color shifts blueward, as expected.  However, once we increase our threshold to remove disks with $b/a\lesssim0.35$, we find an unwanted side effect: increasingly removing the disk population gives increasing weight to bulge--dominated and elliptical populations, yielding a trend toward net redder color.

Second, we investigate a scenario that attempts to avoid this problem.  The lower (dashed--dotted blue) curve shows the effect of removing \emph{any} object regardless of its type (i.e., its $f_{\text{deV}}$ value) as we systematically increase the minimum allowed axis ratio.  We note that at $b/a\gtrsim0.35$ bulge--dominated and elliptical galaxies outnumber disk--dominated galaxies at a ratio of $\sim$3:2; this is the case for the entire volume--limited sample, as well as MW analogs.  Hence, we find that in this case the mean color of the sample becomes increasingly bluer as we increase our minimum allowed $b/a$ threshold above 0.35.  However, since this trend does not stabilize as we push our threshold higher, it is likely that we are increasingly oversampling disk--dominated objects due to preferentially discarding the more prevalent elliptical and bulge--dominated galaxies in this regime.

Lastly, we present an alternative treatment shown by the middle (solid black) curve.  Here, we remove the same disk galaxies that we do in the first scenario, leaving all bulge--dominated and elliptical types initially selected in the sample.  However, we now calculate a weighted mean quantity, where we ensure that the contribution of disk galaxies remaining, after applying any minimum threshold on $b/a$, is equal to that from the disks present in the sample before any cut.  In other words, we calculate the mean property of our filtered sample of galaxies by multiplying the contribution from the remaining disk types after selection by a weighting factor $W=N^\text{before}_\text{disks}/N^\text{after}_\text{disks}$, where $N^\text{before}_\text{disks}$ and $N^\text{after}_\text{disks}$ represent the number of disk galaxies in the sample before and after applying this cut, respectively.  For instance, for $^0(g-r)$ color our estimator reduces to
\begin{align} 
& \langle^0(g-r)\rangle = \nonumber \\
& \frac{N^\text{before}_\text{disks}\,\langle^0(g-r)^\text{after}_\text{disks}\rangle_\text{H--L}+N_\text{ellipticals}\,\langle^0(g-r)_{\text{ellipticals}}\rangle_{\text{H--L}}}{N^\text{before}_\text{disks}+N_\text{ellipticals}},
\label{eq:weighted_mean}
\end{align}
where $N_\text{ellipticals}$ is the number of ellipticals in the MWAS, $\langle^0(g-r)^\text{after}_\text{disks}\rangle_\text{H--L}$ is the mean $^0(g-r)$ color of disk galaxies after the $b/a$ and $f_{\text{deV}}$ cuts are applied, and $\langle^0(g-r)_{\text{ellipticals}}\rangle_\text{H--L}$ is the mean $^0(g-r)$ color of ellipticals in the MWAS.  The H--L subscript here denotes that we are truly using the Hodges--Lehmann estimator of the mean.

By comparing the different curves in Figure \ref{fig:axis_ratio}, it is clear that the weighted mean is favorable over the other scenarios for two reasons.  First, the slope of the weighted--mean curve is shallow enough that moving from a cut of $b/a>0.4$ to $b/a>0.8$ makes a $\lesssim0.01$ magnitude difference in integrated $^0(g-r)$ color.  Hence, compared to the other scenarios, this prescription results in more stable values of mean $^0(g-r)$ over basically all reasonable threshold values of $b/a$.  Second, the position of this curve is between the other two, indicating that we are avoiding giving too much weight to either of the red or blue populations, and hence limiting the impact of any morphological bias.  We note that simply cutting out whole classes of galaxies based on their axis ratios, as in the first two scenarios, provides a more extreme way of dealing with inclination reddening.  The difference measured, however, between the curve for our weighted--mean scenario and either of the other two scenarios is $\lesssim0.02$ mag; all of these differences are significantly less than the statistical errors in $^0(g-r)$ color of the sample ($\sim$0.06 mag).  Based on the results of this exercise, we eliminate any of the original 5000 galaxies selected for the MWAS having both $b/a<0.6$ and $f_{\text{deV}}<0.5$, and we report the reweighted mean property to minimize the impact of any reddening of the sample due to inclination.  As noted above, this cut typically removes $\sim$30\% of a MWAS realization.
\begin{figure*}
\centering
\begin{subfigure}
\centering
\includegraphics[trim=0.4in 0.2in 0.75in 0.8in, clip=true, width=\columnwidth]{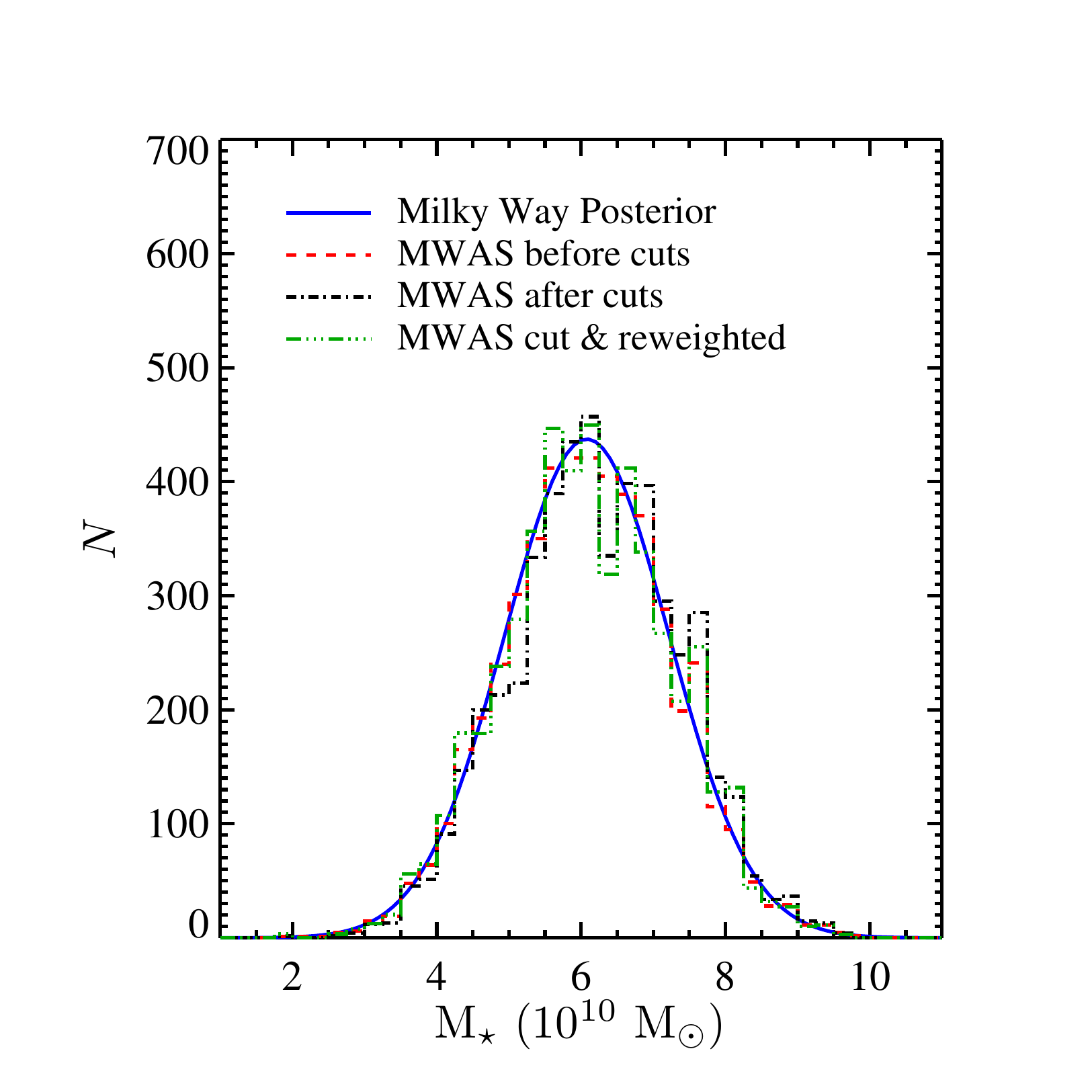}
\end{subfigure}
\begin{subfigure}
\centering
\includegraphics[trim=0.4in 0.2in 0.75in 0.8in, clip=true, width=\columnwidth]{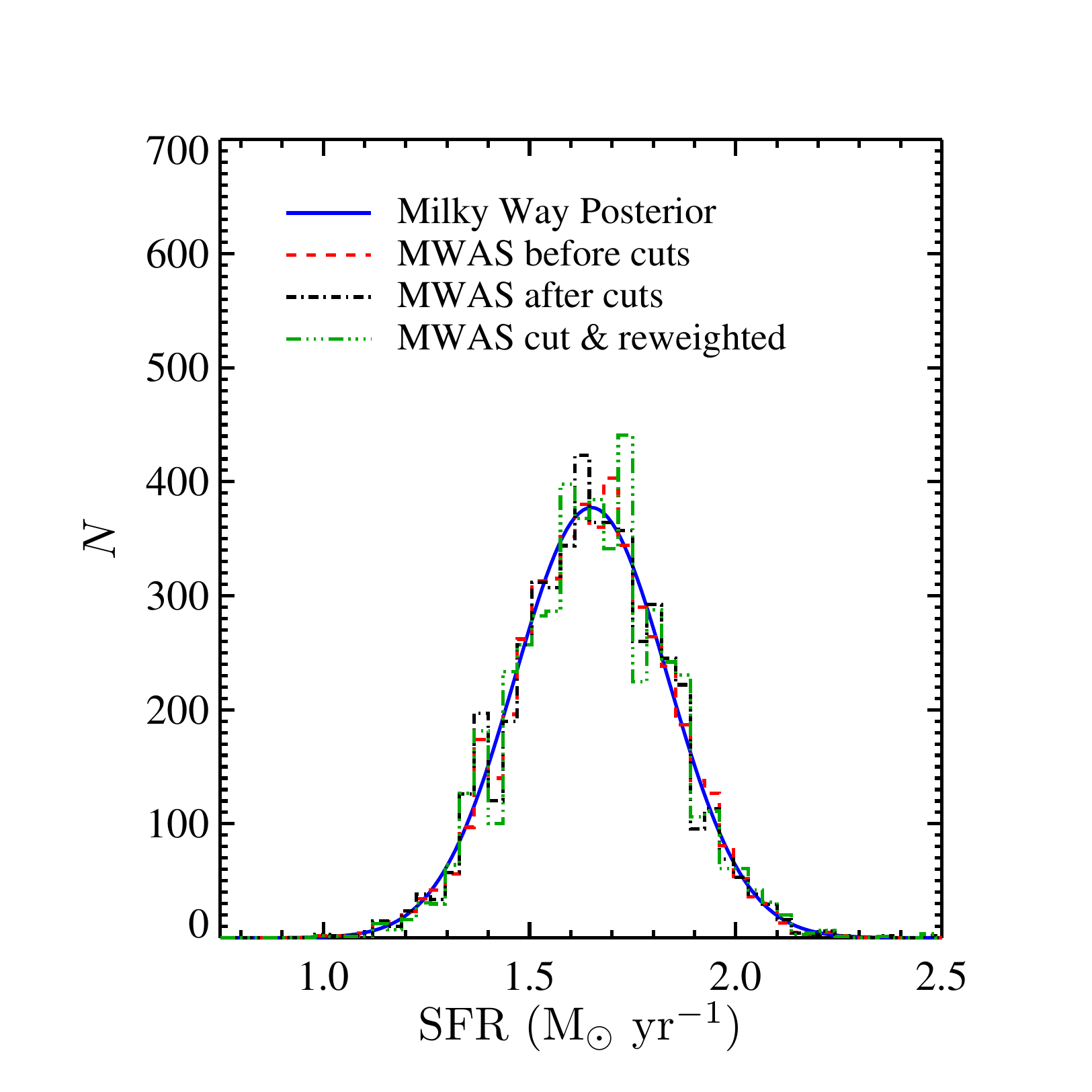}
\end{subfigure}
 \caption{Distribution of stellar masses, \mt\ (\textit{left panel}), and star formation rates (SFRs), \sfr\ (\textit{right panel}), for our Milky Way analog sample at different stages of our analysis procedure overlaid upon the Galactic posterior probability distribution functions (blue solid curves) used for selecting them.  The red dashed line shows the original sample of 5000 analogs drawn before any cuts are applied.  The black dashed--dotted curve shows the remainder of the original sample after removing those that appear to be edge--on disk--dominated systems, whose inclusion would otherwise systematically redden our results for photometric properties, and then renormalizing to reflect the original sample size of 5000.  The green dash--triple--dotted curve shows those objects that make up the black dashed--dotted curve, but reweighted to correct for any morphological bias (i.e., the oversampling of bulge--dominated vs. disk--dominated objects; see Figure \ref{fig:axis_ratio}) that our cuts introduce; see Equation \eqref{eq:weighted_mean}.}
\label{fig:mstar_sfr_hists}
\end{figure*}

We note that if we were to avoid edge--on disks entirely during our selection of MW analogs (they have been selected based only on their \mt\ and \sfr\ values), then our results would suffer from the same morphological bias that is evident from the red dashed curve in Figure \ref{fig:axis_ratio}, but with no way of knowing to what quantitative extent.  Therefore it is important to make a correction only at this stage in the analysis --- i.e., calculating our results via a weighted mean after removing edge--on disks --- so that we may correct for both systematic effects, namely inclination reddening and morphological bias.  Fortunately, we have found that the variations due to such effects are far below the random uncertainties.

One other option would be to correct for inclination--related reddening on an object--by--object basis, rather than trimming and reweighting the sample as was done here.  For instance, \citet{Maller} provide formulae for converting inclination--dependent observed quantities into intrinsic ones.  When applied to the color--magnitude relationship their methods transform the SDSS blue--to--red galaxy ratio from 1:1 to 2:1 in the absolute magnitude range of $-22.75 \le M_K \le -17.75$.  Converting to intrinsic properties could eliminate the need for inclination cuts altogether and allow us to utilize a larger subset of the SDSS main galaxy sample, though at the cost of adopting a particular model for extinction corrections.  In any case, we point out that the difference between our choice of correction and the extreme limits displayed in Figure \ref{fig:axis_ratio} is a $\sim$0.02 mag shift, which turns out to be a factor of 3 times smaller than the uncertainties in our final results.  In addition, the slope of this curve becomes very shallow beyond a minimum axis ratio of 0.4, so any reasonably chosen cut would yield negligible change to our results.  Similar analyses have demonstrated that this same method works well for all colors considered as well as for correcting extinction in absolute magnitudes.  Overall, we expect that any alternative prescription for inclination would have inconsequential impact on the results of this study.

Lastly, in Figure \ref{fig:mstar_sfr_hists} we show the distribution of \mt\ and \sfr\ values for MW analogs compared to the posterior distributions used for selecting them.  This includes distributions for the original sample of 5000 before any cuts, the 3402 galaxies remaining after removing those with $b/a<0.6$ and $f_{\text{deV}}<0.5$, and the reweighted distribution of those 3402 objects in congruence with Equation \eqref{eq:weighted_mean}.  Where necessary, we have renormalized each distribution to reflect a total sample size of 5000 objects.  In all cases, we find that the mean and standard deviation of our sample match those of the posterior distribution and have confirmed that they are Gaussian--distributed via a $Q$--$Q$ plot analysis \citep{Wilk68}.  Hence, we find that our treatment of inclination reddening and morphological bias does not compromise the fundamental design of our MWAS in \sfr--\mt\ space.

\begin{deluxetable*}{ccccc}
\tablewidth{\textwidth}
\tablecaption{ \label{table:props1} Photometric Properties for the Milky Way: Rest--frame $z$=0 SDSS Passbands}
\tablehead{Property & Corrected Value & Bias Removed & $\partial/\partial\mt$ & $\partial/\partial\sfr$ \\
 & (mag) & (mag) & (10$^{-10}$ mag M$_\odot^{-1}$) & (mag M$_\odot^{-1}$ yr)}
\startdata
$^0\!M_u - 5\log h$ & $-19.16_{-0.47}^{+0.57}$ & \phs$0.240\pm0.014$ & $-0.05$ & $-0.32$ \\
$^0\!M_g - 5\log h$ & $-20.36_{-0.41}^{+0.47}$ & \phs$0.142\pm0.011$ & $-0.11$ & $-0.40$ \\
$^0\!M_r - 5\log h$ & $-21.00_{-0.37}^{+0.38}$ & \phs$0.134\pm0.009$ & $-0.11$ & $-0.48$ \\
$^0\!M_i - 5\log h$ & $-21.27_{-0.36}^{+0.38}$ & \phs$0.120\pm0.009$ & $-0.14$ & $-0.49$ \\
$^0\!M_z - 5\log h$ & $-21.56_{-0.37}^{+0.36}$ & \phs$0.126\pm0.009$ & $-0.15$ & $-0.39$ \\
\\[-1ex]
$^0(u-r)$ & \phs$2.043_{-0.157}^{+0.166}$ & \phs$0.090\pm0.0060$ & \phs$0.07$ & $-0.02$ \\
$^0(u-g)$ & \phs$1.358_{-0.093}^{+0.105}$ & \phs$0.077\pm0.0047$ & \phs$0.06$ & \phs$0.02$ \\
$^0(g-r)$ & \phs$0.682_{-0.056}^{+0.066}$ & \phs$0.015\pm0.0017$ & \phs$0.03$ & \phs$0.00$ \\
$^0(r-i)$ & \phs$0.296_{-0.046}^{+0.051}$ & \phs$0.012\pm0.0012$ & \phs$0.01$ & \phs$0.00$ \\
$^0(i-z)$ & \phs$0.291_{-0.041}^{+0.043}$ & $-0.001\pm0.0009$ & \phs$0.01$ & $-0.04$ \\
\enddata
\tablecomments{The Eddington bias estimated for each band, as described in \S\ref{sec:EddBias}, is listed in Column 3.  This is subtracted from the mean property measured from the MWAS, as discussed in \S\ref{sec:inclination} (see Equation \eqref{eq:weighted_mean}), in order to produce the corrected value listed in Column 2.}
\end{deluxetable*} 

\section{Results} \label{sec:results}
With the MWAS assembled and major systematic errors accounted for, we are now able to produce a comprehensive outside--in portrait of our Galaxy.  Table \ref{table:props1} presents the inferred photometric properties we determine for the MW in rest--frame $z$=0 SDSS passbands, and likewise Table \ref{table:props2} presents rest--frame $z$=0.1 SDSS passband results.  The values shown are calculated as the weighted (Hodges--Lehmann estimator of the) mean as described in \S\ref{sec:inclination} and have been corrected for Eddington bias as detailed in \S\ref{sec:EddBias}.  Each row is calculated independently of any other table entry; for instance, we utilize the full distribution of $^0(g-r)$ amongst the MW analogs, rather than deriving this value by subtracting $^0\!M_r$ from $^0\!M_g$ (this is also due to our colors being derived from \texttt{model} magnitudes, whereas absolute magnitudes are based upon \texttt{cmodel}).  For reference we list the inherent Eddington bias that has been subtracted in juxtaposition to each corrected value.

In addition, we tabulate the derivative of each property with respect to total stellar mass and SFR.  This is accomplished by offsetting the distributions we assume for the Galactic \mt\ and \sfr\ by $\pm0.1$ times their respective errors and redoing our analyses.  Along with our fiducial results, this provides three data points to which we fit a quadratic Lagrangian--interpolation polynomial, and then calculate its derivative at the central datapoint.  We choose an offset of $\pm0.1\sigma$ so that the resulting Galactic range in \sfr--\mt\ space does not require selecting a new volume--limited sample of objects.

As discussed in \S\ref{sec:intro}, all extragalactic measurements of $\log\mt$ and $\log\sfr$, measured on the cosmic distance scale, can be converted to reflect different values of the Hubble constant by subtracting from them $2\log(h/0.7)$, effectively shifting them relative to the MW's position in this parameter space.  If we were to instead add this quantity to the Galactic $\log\mt$ and $\log\sfr$ values, the change in our results would be identical; this allows one to calculate how the absolute magnitudes and colors we calculate for the MW change for different $h$ using quantities given in Tables \ref{table:props1}--\ref{table:props4}.  For example,
\begin{align}
\frac{\textrm{d}(^0\!M_r - 5\log h)}{\textrm{d}h} &= \frac{\partial(^0\!M_r - 5\log h)}{\partial\mt} \frac{\textrm{d}\mt}{\textrm{d}h} \nonumber \\
&\quad + \frac{\partial(^0\!M_r - 5\log h)}{\partial\sfr} \frac{\textrm{d}\sfr}{\textrm{d}h}.
\label{eq:deriv}
\end{align}
To be explicit, this means that calculating absolute magnitudes using a different value of $h$ (where $h=H_0/(100 \text{km s}^{-1} \text{Mpc}^{-1})$ has been used) would shift both the positions of the MW and the volume--limited sample together in unison along the absolute--magnitude axis of any CMD we show.  However, calculating \mt\ and \sfr\ values using a different value of $h$ (where $h=0.7$ has been used) would shift the position of the volume--limited sample relative to the MW's position in the CMD; the size of this effect can be estimated using Equation \eqref{eq:deriv}.  For instance, for the MW \mt\ and \sfr\ values we have used along with the values in Table \ref{table:props1}, we find that a $\pm0.05$ shift in $h$ corresponds to a $\sim\pm0.05$ magnitude shift in $^0\!M_r - 5\log h$ and no shift in $^0(g-r)$.  Therefore, we would expect that any reasonable difference between the true value of $h$ and 0.7 will yield negligible changes in our results (well below the measurement uncertainties) and the conclusions we draw from them.
\begin{deluxetable*}{ccccc}
\tablewidth{\textwidth}
\tablecaption{ \label{table:props2} Photometric Properties for the Milky Way: Rest--frame $z$=0.1 SDSS Passbands}
\tablehead{Property & Corrected Value & Bias Removed & $\partial/\partial\mt$ & $\partial/\partial\sfr$ \\
 & (mag) & (mag) & (10$^{-10}$ mag M$_\odot^{-1}$) & (mag M$_\odot^{-1}$ yr)}
\startdata
$^{0.1}\!M_u - 5\log h$ & $-18.85_{-0.51}^{+0.63}$ & $0.271\pm0.012$ & \phs$0.02$ & $-0.38$ \\
$^{0.1}\!M_g - 5\log h$ & $-20.07_{-0.44}^{+0.48}$ & $0.168\pm0.011$ & $-0.10$ & $-0.45$ \\
$^{0.1}\!M_r - 5\log h$ & $-20.78_{-0.39}^{+0.37}$ & $0.130\pm0.009$ & $-0.10$ & $-0.42$ \\
$^{0.1}\!M_i - 5\log h$ & $-21.16_{-0.37}^{+0.38}$ & $0.134\pm0.009$ & $-0.13$ & $-0.59$ \\
$^{0.1}\!M_z - 5\log h$ & $-21.41_{-0.38}^{+0.39}$ & $0.124\pm0.009$ & $-0.14$ & $-0.35$ \\
\\[-1ex]
$^{0.1}(u-r)$ & \phs$2.201_{-0.172}^{+0.201}$ & $0.105\pm0.0072$ & \phs$0.10$ & \phs$0.08$ \\
$^{0.1}(u-g)$ & \phs$1.419_{-0.112}^{+0.124}$ & $0.074\pm0.0052$ & \phs$0.07$ & \phs$0.04$ \\
$^{0.1}(g-r)$ & \phs$0.782_{-0.063}^{+0.081}$ & $0.031\pm0.0027$ & \phs$0.02$ & $-0.03$ \\
$^{0.1}(r-i)$ & \phs$0.390_{-0.042}^{+0.046}$ & $0.001\pm0.0010$ & \phs$0.01$ & $-0.01$ \\
$^{0.1}(i-z)$ & \phs$0.275_{-0.046}^{+0.047}$ & $0.010\pm0.0011$ & \phs$0.01$ & $-0.05$ \\
\enddata
\tablecomments{The Eddington bias estimated for each band, as described in \S\ref{sec:EddBias}, is listed in Column 3.  This is subtracted from the mean property measured from the MWAS, as discussed in \S\ref{sec:inclination} (see Equation \eqref{eq:weighted_mean}), in order to produce the corrected value listed in Column 2.}
\end{deluxetable*} 

Similarly, the results for the total stellar mass of the MW found in LN15 would be changed if any adjustments are made to the absolute distance scale (see Table 6 of that paper).  Predominantly this manifests in changes to the Galactocentric radius of the Sun, $R_0$; LN15 conservatively used $8.33\pm0.35$ kpc based on the work of \citet{Gillessen}.  Firstly, since we found that $\partial\mt/\partial R_0=3.09\times10^{10}$ \massunits\ kpc$^{-1}$ for the MW, the impact of a change in $R_0$ can be obtained by replacing $h$ with $R_0$ in Equation \eqref{eq:deriv} (note $\partial\sfr/\partial R_0=0$).

Uncertainties in $R_0$ dominate the error budget in our \mt\ model.  We find that if we were to instead adopt $R_0=8.36\pm0.11$ kpc based on \citet{Chatzo}, yielding a $\sim$69\% decrease in uncertainty in $R_0$, then the total stellar mass from LN15 becomes $\mt=6.18\pm0.50\times10^{10}$ \massunits, corresponding to a net $\sim$57\% decrease in \mt\ uncertainty.  This ultimately yields a $\sim$20\% decrease in $M_r$ uncertainty, while also causing the MW analogs to lie along a tighter trend in color--magnitude space.  Ultimately, as our knowledge of the structure of our Galaxy improves (e.g., by measurements from Gaia), our methods should be able to more strongly constrain the MW's location in CMDs.  In contrast, the current uncertainties in the Galactic SFR have a negligible effect on the constraints on the photometric properties derived in this paper.  This is because the uncertainty in the MW stellar mass is a significant fraction of the range of stellar masses amongst galaxies of comparable SFR, while the SFR uncertainty is a $\sim$7$\times$ smaller fraction of the range of SFRs at fixed mass.  We note also that the evolution of galaxies since $z\sim0.1$ appears to have negligible effect on our results; e.g., limiting our analysis instead to objects at $0.045 < z < 0.075$ yields differences in our results that are much smaller than the errors.

\begin{figure}
\centering
\includegraphics[trim=0.5in 0.2in 0.85in 0.8in, clip=true, width=\columnwidth]{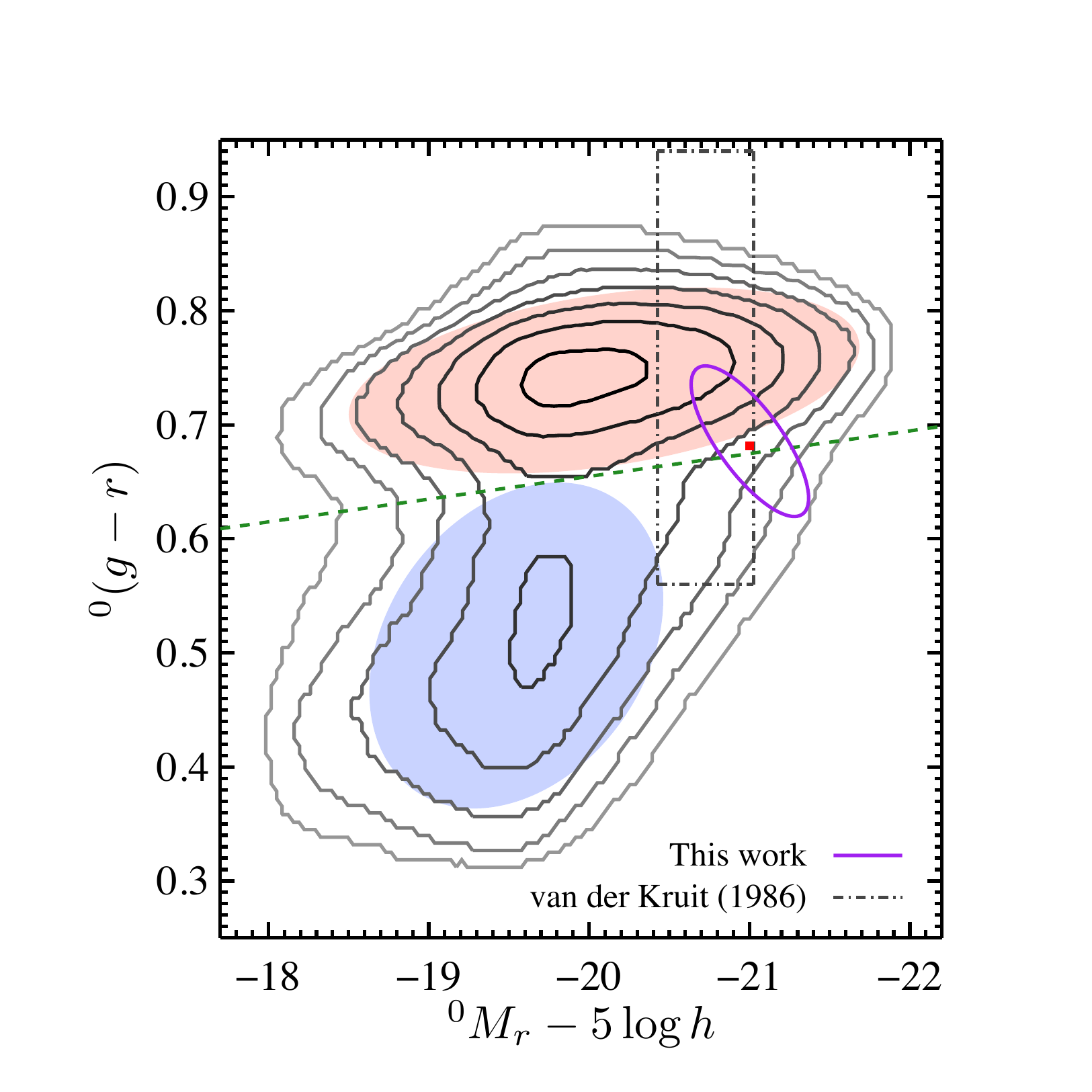}
\caption{Eddington--bias--corrected position of the Milky Way in SDSS $^0(g-r)$ vs. $^0\!M_r$ color--magnitude space (red point and purple 1$\sigma$ ellipse).  For comparison, we show in gray dashed--dotted lines the previously best 1$\sigma$ constraint directly measured by vdK86, converted to the SDSS AB magnitude system via transformation equations from \citet{Cook}.  In order to place this measurement in this plot we subtract $5\log(h/0.7)$, allowing it to be directly compared to the SDSS sample.  Log--spaced contours show the density of galaxies in our volume--limited sample; we shade the core of the red--sequence and blue--cloud regions in red and blue, respectively, and show the same green dashed color division line as in Figure \ref{fig:MW_analogs_cmd}.  Until now, the Milky Way's position has remained highly uncertain in this parameter space.  Our new measurement dramatically improves our knowledge of how the Galaxy compares to others in the local Universe; we likely straddle the division between the blue--cloud and red--sequence populations, or the so--called ``green--valley'' region of this diagram.  This ranks the Milky Way amongst the brightest and reddest spiral galaxies still producing new stars today.  It may well be in a transitional evolutionary phase where star formation is dying out.}
\label{fig:MW_ellipse_cmd}
\end{figure}

Figure \ref{fig:MW_ellipse_cmd} now shows the position of the MW corrected for Eddington and inclination bias, as listed in Table \ref{table:props1}, as a red dot in rest--frame SDSS $^0(g-r)$ vs. $^0\!M_r$ space; it is overlaid upon log--spaced density contours for the volume--limited sample.  The purple ellipse displays our 1$\sigma$ confidence region, accounting for the covariance between color and absolute magnitude; this yields a vast improvement in constraining how our Galaxy fits amongst the extragalactic population compared to the previously best 1$\sigma$ constraints from \citet[][gray dashed--dotted lines]{vanderKruit}.  For convenience, we have highlighted the red--sequence and blue--cloud regions of this diagram (flipped in position compared to \sfr--\mt\ space, since higher--SFR galaxies are bluer).  We see that MW's position straddles the boundary between these two populations, with a chance that it lies in the core of the red sequence or redder.  In addition, given that the blue cloud includes the vast majority of the spiral galaxy population (\citealp{Strateva01}; \citealp{Blanton03}; \citealp{Wong12}; S14), we see that our value of $^0\!M_r - 5\log h=-21.00_{-0.37}^{+0.38}$ establishes the MW amongst the brightest spiral galaxies in the local universe, while its integrated color of $^0(g-r)=0.682_{-0.056}^{+0.066}$ ranks it amongst the reddest as well.

Lastly, we produce an updated plot equivalent to Figure 1 of M11 by showing our constraints on the MW's position in $^0(u-r)$ vs. \mt\ space, where the green valley becomes stretched out and more distinguishable.  Here, we have highlighted the green--valley region based on two different definitions.  First, the dark green region shows the division line empirically derived for SDSS galaxies by \citet{Baldry} with an offset of $\pm$0.1 mag in the vertical direction (the definition of the green valley employed by M11), which matches well with the density contours for our volume--limited sample.  Second, the light green region shows a definition based upon correcting all SDSS galaxies for dust effects, as defined by \citet[][hereafter S14]{Schawinski}.  In the second case, many of the intermediate--color objects are blue galaxies that are both dusty and viewed edge--on, and so switching to intrinsic (face--on) properties moves this population blueward in the plot, effectively thinning out and expanding the green--valley region more.  Given that our measurement of the MW's position in this space is effectively face--on, the green--valley definition from S14 provides a suitable comparison.
\begin{deluxetable*}{ccccc}
\tablewidth{\textwidth}
\tablecaption{ \label{table:props3} Photometric Properties for the Milky Way: Rest--frame $z$=0 Johnson--Cousins Passbands}
\tablehead{Property & Corrected Value & Bias Removed & $\partial/\partial\mt$ & $\partial/\partial\sfr$ \\
 & (mag) & (mag) & (10$^{-10}$ mag M$_\odot^{-1}$) & (mag M$_\odot^{-1}$ yr)}
\startdata
$^0\!M_U - 5\log h$ & $-20.02_{-0.47}^{+0.50}$ & $0.232\pm0.011$ & $-0.10$ & $-0.61$ \\
$^0\!M_B - 5\log h$ & $-20.07_{-0.44}^{+0.40}$ & $0.173\pm0.009$ & $-0.12$ & $-0.49$ \\
$^0\!M_V - 5\log h$ & $-20.74_{-0.39}^{+0.37}$ & $0.132\pm0.008$ & $-0.12$ & $-0.37$ \\
$^0\!M_R - 5\log h$ & $-21.26_{-0.36}^{+0.40}$ & $0.131\pm0.007$ & $-0.11$ & $-0.16$ \\
$^0\!M_I - 5\log h$ & $-21.84_{-0.39}^{+0.36}$ & $0.125\pm0.008$ & $-0.12$ & $-0.34$ \\
\\[-1ex]
$^0(U-V)$ & \phs$0.890_{-0.123}^{+0.148}$ & $0.094\pm0.0055$ & \phs$0.02$ & $-0.14$ \\
$^0(U-B)$ & \phs$0.149_{-0.070}^{+0.078}$ & $0.063\pm0.0038$ & \phs$0.01$ & $-0.04$ \\
$^0(B-V)$ & \phs$0.744_{-0.054}^{+0.068}$ & $0.028\pm0.0022$ & \phs$0.01$ & $-0.05$ \\
$^0(V-R)$ & \phs$0.541_{-0.042}^{+0.046}$ & $0.005\pm0.0008$ & \phs$0.00$ & $-0.02$ \\
$^0(R-I)$ & \phs$0.598_{-0.049}^{+0.047}$ & $0.007\pm0.0009$ & \phs$0.01$ & \phs$0.01$ \\
\enddata
\tablecomments{Values in this table are determined from analyzing the distributions of properties for Milky Way analogs, but after transforming SDSS $ugriz$ measurements to Johnson--Cousins $UBVRI$--equivalent values on an object--by--object basis using the \texttt{kcorrect} software.  As a reminder, $UBVRI$ magnitudes are on the Vega system, whereas $ugriz$ magnitudes are on the AB system.}
\end{deluxetable*}

Compared to the prior constraints (gray dashed--dotted lines), we are in a much better position to now identify where the MW lies relative to other galaxies.  In particular, our Galaxy appears to be entering, if not already a part of, the green--valley region where objects are expected to be in a transitional phase; here star formation is quenching by some mechanism(s); consequentially, green--valley galaxies are moving on a trajectory toward the red sequence (for more detail see, e.g., \citealt{Gon}; \citealt{Fang}; S14).

In Tables \ref{table:props3} and \ref{table:props4} we present our results for MW properties transformed to the Johnson--Cousins passband system.  As a reminder, these values have been calculated in an identical fashion to those listed in Tables \ref{table:props1} and \ref{table:props2}, but after transforming each MW analog's set of SDSS $ugriz$ magnitudes to an equivalent set of Johnson--Cousins $UBVRI$ measurements using the \texttt{kcorrect} software package \citep{kcorrect}.  This entails calculating magnitudes in $UBVRI$ passbands from the linear combination of template galaxy SEDs from BC03 that best fits the observed SDSS $ugriz$ photometry on an object--by--object basis, and hence should provide the most accurate transformations.  A viable alternative would be to apply the empirical color transformations provided by \citet{Cook} directly to our results in Tables \ref{table:props1} and \ref{table:props2}, though these equations represent the mean transformations between the two passband systems averaged over galaxies with a range of morphologies, SFRs, etc.  Nevertheless, we find that applying the \citealt{Cook} transformation equations to our SDSS results produces estimates on the Johnson--Cousins system that are quite similar to our nominal values determined using \texttt{kcorrect}.  The differences are almost always at the 0.1--0.3$\sigma$ level (including for mass--to--light ratios, which we discuss next), the one exception being $^0(U-B)$, where the two methods agree at the 0.75$\sigma$ level.  Note that we have used the \citeauthor{Cook} equations to transform the \citet{vanderKruit} result in Figure \ref{fig:MW_ellipse_cmd}.
\begin{deluxetable*}{ccccc}
\tablewidth{\textwidth}
\tablecaption{ \label{table:props4} Photometric Properties for the Milky Way: Rest--frame $z$=0.1 Johnson--Cousins Passbands}
\tablehead{Property & Corrected Value & Bias Removed & $\partial/\partial\mt$ & $\partial/\partial\sfr$ \\
 & (mag) & (mag) & (10$^{-10}$ mag M$_\odot^{-1}$) & (mag M$_\odot^{-1}$ yr)}
\startdata
$^{0.1}\!M_U - 5\log h$ & $-20.10_{-0.51}^{+0.60}$ & $0.252\pm0.011$ & $-0.11$ & $-0.64$ \\
$^{0.1}\!M_B - 5\log h$ & $-19.96_{-0.45}^{+0.49}$ & $0.196\pm0.010$ & $-0.12$ & $-0.56$ \\
$^{0.1}\!M_V - 5\log h$ & $-20.47_{-0.40}^{+0.41}$ & $0.136\pm0.009$ & $-0.12$ & $-0.33$ \\
$^{0.1}\!M_R - 5\log h$ & $-20.98_{-0.35}^{+0.46}$ & $0.143\pm0.008$ & $-0.12$ & $-0.25$ \\
$^{0.1}\!M_I - 5\log h$ & $-21.60_{-0.37}^{+0.41}$ & $0.139\pm0.008$ & $-0.13$ & $-0.36$ \\
\\[-1ex]
$^{0.1}(U-V)$ & \phs$0.604_{-0.135}^{+0.159}$ & $0.099\pm0.0059$ & \phs$0.02$ & $-0.10$ \\
$^{0.1}(U-B)$ & $-0.014_{-0.090}^{+0.096}$ & $0.055\pm0.0034$ & \phs$0.01$ & $-0.01$ \\
$^{0.1}(B-V)$ & \phs$0.626_{-0.062}^{+0.073}$ & $0.037\pm0.0031$ & \phs$0.01$ & $-0.09$ \\
$^{0.1}(V-R)$ & \phs$0.518_{-0.043}^{+0.049}$ & $0.010\pm0.0010$ & \phs$0.00$ & $-0.04$ \\
$^{0.1}(R-I)$ & \phs$0.637_{-0.047}^{+0.048}$ & $0.006\pm0.0009$ & \phs$0.01$ & \phs$0.01$ \\
\enddata
\tablecomments{Values in this table are determined from analyzing the distributions of properties for Milky Way analogs, but after transforming SDSS $ugriz$ measurements to $UBVRI$--equivalent values on an object--by--object basis using the \texttt{kcorrect} software.  As a reminder, $UBVRI$ magnitudes are on the Vega system, whereas $ugriz$ magnitudes are on the AB system.}
\end{deluxetable*}

In addition to the photometric properties presented in Tables \ref{table:props1}--\ref{table:props4}, we also provide in Table \ref{table:M2L} new estimates of global stellar mass--to--light ratios, $\Upsilon^\star$, for the MW for all SDSS and Johnson--Cousins passbands in the $z$=0 and $z$=0.1 rest frames.  These are calculated from the full distribution of $\Upsilon^\star$ values for the MWAS, in the same manner as we calculate photometric properties.  To do so, we first calculate the stellar mass--to--light ratio for each MW analog in passband $x$ in the rest frame of redshift $z$ as
\begin{equation}
^z\Upsilon^\star_x = \mt\times10^{0.4((^z\!M_x + 5\log(0.7/h) - ^z\!M_{x,\odot})}\,L_\odot^{-1}, \label{eq:m2l}
\end{equation}
where $^z\!M_{x,\odot}$ and $L_\odot$ represent the absolute magnitude and luminosity, respectively, of the Sun, which we calculate using the \texttt{k\_solar\_magnitudes} routine from the \texttt{kcorrect} package.  We note that Equation \eqref{eq:m2l} is written to make it clear that we have converted absolute magnitudes to reflect $h=0.7$ and hence be on the same scale as our \mt\ values; however, it should be noted that $\Upsilon^\star$ is intrinsically a cosmology--independent quantity.  For instance, if we now chose to rescale quantities from $h=0.7$ to 0.8, the right hand side of Equation \eqref{eq:m2l} would gain a factor of $(0.7/0.8)^2$ for the change in stellar mass and a factor of $10^{0.4(5\log(0.8/0.7))}$ for the change in luminosity, which cancel.  Next, we use Equation \eqref{eq:weighted_mean}, replacing $^0(g-r)$ with $^z\Upsilon^\star_x$, in order to produce our weighted--mean estimate.  Lastly, we multiply this by a factor of $10^{-0.4B}$, where $B$ is the Eddington bias correction, which is listed in Tables \ref{table:props1}--\ref{table:props4}.  In this way, our results are corrected for Eddington bias, inclination effects, and morphological bias (all subdominant to the random errors), consistent with all other properties presented here.

\begin{deluxetable}{lccccc}
\tablewidth{\columnwidth}
\tablecaption{ \label{table:M2L} Global Stellar Mass--to--light Ratios for the Milky Way }
\tablehead{Rest-- & \multirow{2}{*}{$\Upsilon^\star_u$} & \multirow{2}{*}{$\Upsilon^\star_g$} & \multirow{2}{*}{$\Upsilon^\star_r$} & \multirow{2}{*}{$\Upsilon^\star_i$} & \multirow{2}{*}{$\Upsilon^\star_z$} \\
frame &&&&&}
\startdata
$z$=0 & $1.90^{+1.18}_{-0.80}$ & $1.96^{+0.69}_{-0.64}$ & $1.66^{+0.63}_{-0.49}$ & $1.43^{+0.48}_{-0.41}$ & $1.11^{+0.32}_{-0.32}$ \\
\\
$z$=0.1 & $1.77^{+1.61}_{-0.83}$ & $1.93^{+0.81}_{-0.68}$ & $1.84^{+0.64}_{-0.57}$ & $1.54^{+0.61}_{-0.44}$ & $1.26^{+0.38}_{-0.37}$ \\ \\
\hline
\hline
\\[-1.4ex]
Rest-- & \multirow{2}{*}{$\Upsilon^\star_U$} & \multirow{2}{*}{$\Upsilon^\star_B$} & \multirow{2}{*}{$\Upsilon^\star_V$} & \multirow{2}{*}{$\Upsilon^\star_R$} & \multirow{2}{*}{$\Upsilon^\star_I$} \\
frame &&&&& \\
\hline
\\[-1.4ex]
$z$=0 & $1.86^{+1.05}_{-0.80}$ & $1.89^{+0.78}_{-0.65}$ & $1.86^{+0.69}_{-0.58}$ & $1.61^{+0.59}_{-0.48}$ & $1.29^{+0.43}_{-0.37}$ \\
\\
$z$=0.1 & $1.81^{+1.39}_{-0.84}$ & $1.85^{+0.97}_{-0.68}$ & $1.94^{+0.82}_{-0.62}$ & $1.74^{+0.59}_{-0.53}$ & $1.43^{+0.47}_{-0.42}$ \\
\enddata
\tablecomments{\footnotesize See the end of \S\ref{sec:results} for details on the calculation of these values, which are expressed in units of $\massunits/L_\odot$.}
\end{deluxetable} 

\section{Summary and Discussion} \label{sec:summary}
This paper has focused on determining the global photometric properties of the MW to facilitate comparisons to observations of other galaxies.  In LN15 we have derived a new, highly constrained stellar mass and SFR for the MW using a multitude of independent results from the literature, which encompass many different methods and datasets.  We then identified a set of SDSS galaxies analogous to the MW, whose distribution of SFR and total stellar mass values match the probability distributions for these quantities (given uncertainties) of the MW.  These two quantities are strongly correlated with a galaxy's luminosity and color \citep[see, e.g.,][]{BelldeJong}, so a galaxy that matches our Galaxy in stellar mass and SFR would also be expected to have a similar overall SED.  We then determine the range of photometric properties of these galaxies, allowing us to constrain MW properties in a manner that is largely robust to the effects of Galactic extinction (unless the MW is so unusual that it has no true analogs amongst the set of galaxies matching its \mt\ and \sfr).  We have accounted for the Eddington bias involved with selecting galaxies based on their SFR and stellar mass, and tested the impact of reddening effects on this sample.  In \S\ref{sec:results} we have provided a full tabulation of useful MW photometric properties.

\subsection{Comparisons to Earlier Color Measurements} \label{sec:comp_color}
Overall, the results from our MW analog--based analysis method compare well with literature estimates of the properties of the MW.  Since many of those estimates are made using the Johnson--Cousins passband system, we will often rely on our results transformed to this system in order to make direct comparisons; these are available in Tables \ref{table:props3}--\ref{table:M2L}.  First and foremost, our transformed estimate of $^0(B-V) =0.744_{-0.054}^{+0.068}$ is in excellent agreement with the widely used vdK86 measurement of $0.83\pm0.15$, consistent at the $\sim$0.5$\sigma$ level.  Our result indicates a slightly bluer color for the MW with a smaller uncertainty by a factor of $\sim$3.

As mentioned in \S\ref{sec:intro}, earlier measurements yielded much bluer color estimates for the MW than vdK86, and hence also much bluer than the estimate we have presented here.  The dV\&P two--component model produced a color estimate of $B-V=0.53\pm0.05$, which is inconsistent at nearly the 3$\sigma$ level with our estimate.  The B\&S two--component model yields $B-V=0.45$; given the lack of any error estimates, this is difficult to compare to our value, though again significantly bluer.  \citet{Bahcall86} advised using a $\pm0.2$ mag margin of error when comparing colors to the model, given the wide variety of systematic uncertainties existing in the data at that time.  If we use this as the error estimate for the B\&S model value, we find that our result is redder by $0.29\pm0.21$ magnitudes, making these estimates inconsistent at the $\sim$1.4$\sigma$ significance level.  It is possible that the tension between the dV\&P and B\&S color estimates and the one presented here would be reduced if the two--component models employed were updated to more current constraints on the Galaxy's stellar populations.

In dV83, an estimated color of $B-V=0.53\pm0.04$ is quoted for the MW, obtained by averaging the observed colors of nearby Sb/c types. This value appears to originate from data in Table 4 of \citet[][hereafter dV77]{dV1977}, which indicates that the distribution of corrected colors for a sample of 70 Sbc galaxies is described by $B-V=0.564\pm0.066$.  The measurements for each object are tabulated in the \textit{Second Reference Catalogue of Bright Galaxies} \citep[RC2;][]{RC2}, which collected extragalactic data published since the 1930s.  Each $B-V$ color measurement in RC2 is corrected to the asymptotic total light from each galaxy using a Laplace--Gauss integral technique, as a function of its morphological type ($T$) and the effective aperture diameter ($A_e$) containing 50\% of its total light (in some cases, the $B-V$ color is transformed from measurements in different passbands).  Each ``total'' $B-V$ color is then corrected to zero Galactic extinction via a model of the Galactic dust as a function of coordinates ($l,b$), to zero internal extinction based on a model of inclination reddening as a function of $T$ and isophotal axis ratio, and to the $z=0$ rest frame via a $K$--correction modeled as a function of $T$ and $z$.

Given the difficulties of these corrections, as well as the challenges of properly intercalibrating photographic and photoelectric measurements from a wide variety of sources, it is likely that there could be significant systematic errors in this mean $B-V$ estimate.  Furthermore, dV83 assumes $T=4$ for the MW (no less than 2.5 and no more than 5.5) and quotes the rate of change of the mean corrected color along the $T$ sequence near $T=4$ to be -0.10.  While $T=4$ (or equivalently Sb/c) fits well with the Galactic bulge--to--total ratio of 0.15 we have found in LN15, the uncertainty in the MW's morphological type will still represent an additional source of uncertainty that appears not to have been included in the error estimate from dV83.  We can therefore only treat the uncertainties quoted in the dV83 measurement as a lower limit.  If we instead consider the value of $B-V=0.564\pm0.066$ from dV77, this is bluer than our nominal result by 0.18 mag and inconsistent at the $\sim$2$\sigma$ significance level.  We believe the tension between our color measurement (or that of vdK86) and the estimates from dV77 and dV83 would be relieved if the sources of uncertainty described above were included in this estimate.

For comparison, \citet{Fukugita95} performed a similar analysis for galaxy types across the Hubble sequence by comparing synthetic colors measured from galaxy SEDs to broadband photometry taken from the \textit{Third Reference Catalogue of Bright Galaxies} \citep{RC3}.  Listed in their Table 2, the authors found that the average $B-V$ color for 676 Sb/c types (using only objects with $|b|>30\,^{\circ}$, but applying no reddening correction) is $0.68\pm0.14$, which is in excellent agreement with our MW result.  More recently, \citet{Lorenzo2012} investigated the colors of isolated galaxies in the AMIGA sample that are also found in SDSS--DR8.  This sample included 466 galaxies, two--thirds of which were classified as Sb/c.  Similarly to the methods we employ, the authors used \texttt{model} magnitudes that were corrected for Galactic dust extinction and $K$--corrected to $z$=0 rest--frame passbands.  Listed in their Table 3, they found that the median $^0(g-r)$ color for Sb/c types is $0.65\pm0.09$, which compares well with our MW $^0(g-r)$ estimate of $0.682_{-0.056}^{+0.066}$.  That table also provides colors for a variety of Sb/c galaxy samples; these subsets vary in local environment and redshift range, but all yield color estimates that agree with our MW value at or below the $\sim$1$\sigma$ significance threshold.

\subsection{Comparisons to Earlier Absolute Magnitude Measurements} \label{sec:comp_absmag}
Comparisons of absolute magnitudes require more care, as they require additional assumptions that are prone to systematic error, particularly the value of $h=H_0/(100 \text{km s}^{-1} \text{Mpc}^{-1})$ used to bring extragalactic distance estimates (determined from $z$) and measurements based on absolute distances (in pc) onto a common scale.  For the following discussion we adopt $h=0.7$.  The vdK86 study yielded estimates of $M_B=-20.3\pm0.2$ and (when combined with his $B-V$ estimate) $M_V=-21.1\pm0.3$; these compare well with our slightly brighter estimates of $^0\!M_B=-20.84_{-0.44}^{+0.40}$ and $^0\!M_V=-21.51_{-0.39}^{+0.37}$, which are consistent at the $\sim$1$\sigma$ level.  The B\&S two--component model yields $M_B=-20.1$ and $M_V=-20.5$, measurably dimmer than the results we have found, though again hard to compare to with no error estimates given.  The dV\&P two--component model, on the other hand, produced $M_B=-20.2\pm0.15$ (dV83) and $B-V=0.53\pm0.05$, leading to $M_V=-20.7\pm0.16$; these values are inconsistent with our $^0\!M_B$ and $^0\!M_V$ results at the $\sim$1.5$\sigma$ and $\sim$2$\sigma$ levels, respectively.  

More recently, \citet{Flynn} analyzed Hipparcos and Tycho data for the local disk and extrapolated using an exponential disk model (in combination with earlier bulge luminosity estimates) to determine $M_I = -22.3\pm0.17$.  This compares well with our brighter value of $-22.61_{-0.39}^{+0.36}$, and is consistent with it at the $\sim$0.8$\sigma$ level.  Also, \citet{Liu} converted the best--to--date Vega--calibrated $M_V$ measurement for the MW \citep{vandenBergh} into an AB--calibrated absolute $^{0.1}r$--band magnitude of -21.97 (with no error estimate given); this is within $\sim$1$\sigma$ of our estimate of $^{0.1}\!M_r=-21.55_{-0.39}^{+0.37}$, but brighter, rather than fainter.

\subsection{Comparisons to Luminosity Function Measurements} \label{sec:comp_lumfunc}
\citet{BlantonL*} determined luminosity functions for galaxies in all SDSS passbands using the SDSS Early Data Release.  These were determined as the Schechter function that fits best to the distribution of Petrosian absolute magnitudes of galaxies, converted to the AB system, $K$--corrected to $z=0.1$ rest--frame $ugriz$ passbands, and corrected for galaxy luminosity evolution; this should compare well with the \texttt{cmodel} absolute magnitudes used in this work after correcting them for the luminosity evolution since $z=0.1$.  The Schechter function is parameterized by the characteristic absolute magnitude, $M_*$ (not to be confused with the total stellar mass which we have denoted as \mt), which provides a measure of where the luminosity function transitions from being well fit by a power law into an exponential drop--off.  Thus, galaxies with increasing absolute magnitude beyond $M_*$ rapidly become more rare.  To compare our results, we add $0.1Q$ to the results listed in Table \ref{table:props2} of this paper, where $Q$ is the appropriate correction in units of magnitude per unit redshift for each band as listed in Table 3 of \citet{BlantonL*}, and then subtract from this quantity the appropriate $M_*$ value for each band as listed in their Table 2.  Based on this work, we find that the the MW is brighter than their $M_*$ by $0.50\pm0.64$, $0.48\pm0.48$, $0.18\pm0.37$, $0.18\pm0.38$, and $0.15\pm0.39$ magnitudes in the $^{0.1}ugriz$ bands, respectively, essentially showing the MW to be consistent with $M_*$ in all bands at the $\lesssim$1$\sigma$ significance level.  Similarly, \citet{Dorta} reproduced the analysis of \citet{BlantonL*} using SDSS Data Release 6, which provides larger redshift--complete samples of galaxies and incorporates improved reductions of SDSS imaging data.  However, the luminosity functions that result from this work neglect any correction for the evolution of galaxies, as its impact is estimated to be very small for the redshift ranges used (i.e., $z\lesssim0.2$).  To compare our results with this work, we subtract $M_*$ listed for the appropriate band in their Table 2 from our values in Table \ref{table:props2} of this paper.  Based on this work, we find that the MW is brighter than $M_*$ by $1.13\pm0.63$, $0.54\pm0.48$, $0.07\pm0.37$, $0.23\pm0.38$, and $0.01\pm0.39$ magnitudes in the $^{0.1}ugriz$ bands, respectively.  Again, we find the MW to be consistent with their $M_*$ in nearly all bands at the $\sim$1$\sigma$ confidence level, and hence is comparable in luminosity to $L_*$ galaxies in the nearby universe.

\subsection{Comparisons to Green--valley Definitions} \label{sec:comp_greenvalley}
\begin{figure}
\centering
\includegraphics[trim=0.5in 0.2in 0.65in 0.8in, clip=true, width=\columnwidth]{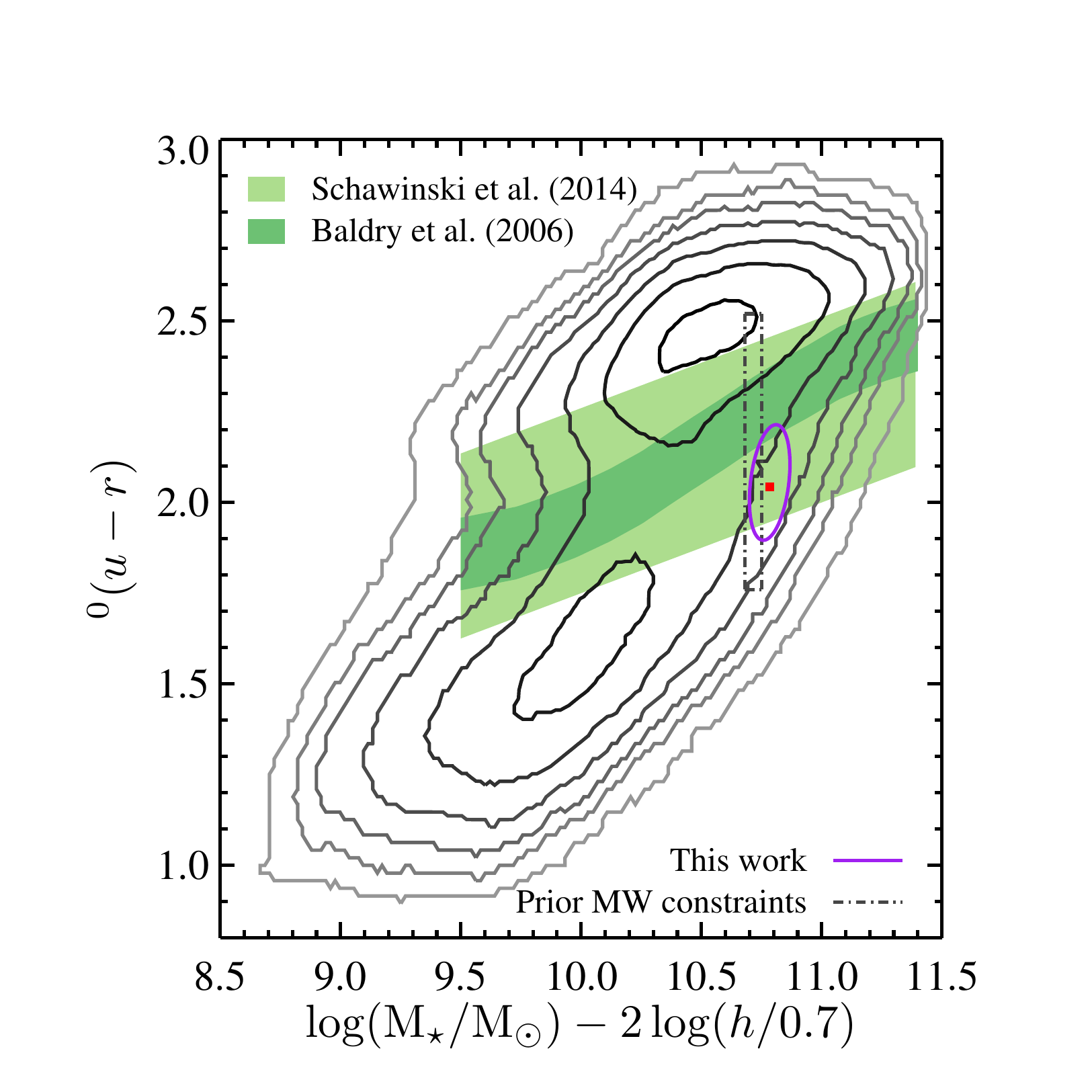}
\caption{Updated version of Figure 1 from M11 showing the Milky Way's corrected position (red point and purple 1$\sigma$ ellipse) in $^0(u-r)$ vs. \mt\ space, where again our new constraints are a dramatic improvement upon and consistent with the prior measurements (gray dashed--dotted lines) utilized by M11 (updated here to the color transformations for galaxies from \citealt{Cook}).  Comparing with Figure \ref{fig:MW_ellipse_cmd}, the green valley becomes much stretched out in $^0(u-r)$ color space.  The dark shaded green region follows the same prescription as M11, using the empirically derived \citet{Baldry} color division line with a $\pm$0.1 $^0(u-r)$ offset.  Second, the light shaded green region is the green valley as defined by S14 for SDSS galaxies after the effects of dust are removed; this provides a suitable comparison for our dust--corrected Milky Way results, whereas the grayscale contours do not reflect this correction.  A similar story emerges as from Figure \ref{fig:MW_ellipse_cmd}: our Galaxy likely resides in the saddle of the bimodal color distribution of galaxies in the local universe.  Measured externally, it would appear redder than the majority of spiral galaxies, yet bluer than most ellipticals.  This makes the Milky Way one of most massive, brightest, and reddest of spiral galaxies with appreciable star formation today.}
\label{fig:MW_mutch_cmd}
\end{figure}

In Figure \ref{fig:MW_ellipse_cmd}, we have presented the MW's location in the $^0(g-r)$ vs. ${^0M}_r$ plane, demonstrating that it falls in the intermediate region between the blue--cloud and red--sequence populations.  Our results are consistent with the hypothesis posed by M11 that the MW could be a member of the ``green--valley'' population.  In Figure \ref{fig:MW_mutch_cmd}, we have produced an updated $^0(u-r)$ vs. $\log\mt$ diagram modeled on of Figure 1 of M11, showing the vast improvement in our constraints compared to those from prior measurements.  Here, the MW lies bluer than their definition of the green valley, i.e., the region within 0.1 mag of the SDSS color division line determined by \citet{Baldry}.

However, more recent work by S14, using $K$--corrected and dust--corrected DR7 magnitudes for galaxies at $0.02 < z < 0.05$ (comparable to our sample at $0.03<z<0.09$), defines the green valley to be $-0.75< {^0(u-r)} - 0.25\log(\mt/\massunits)<-0.24$; this definition would indicate that the MW is in fact a green--valley galaxy in this diagram.  \citet{Jin} present a definition of the green valley that avoids dust--reddening effects by using face--on nearby galaxies with DR7 magnitudes $K$--corrected to $z=0.1$.  They define the center of the valley to be $^{0.1}(u-r)=-0.121(^{0.1}\!M_r - 5\log h) - 0.061$ (with no range given).  Given the uncertainties in our measurements, we find that the MW is bluer than this line by $0.25\pm0.21$ mag, consistent with it at the $\sim$1.2$\sigma$ confidence level.  

\citet{Mendez11} define the green--valley region of the $^0(U-B)$ vs. $^0\!M_B$ plane for AEGIS galaxies to be within a $\pm$0.1 mag vertical offset of the line $^0(U-B)=-0.0189(^0\!M_B - 5\log h) - 0.32$ (where we have converted from AB to Vega magnitudes).  We find that our results place the MW redder than this line by $0.089\pm0.070$ mag; hence the MW might be considered a green--valley galaxy by this definition.  \citet{Willmer06} present a similar CMD division line for red and blue galaxies measured in the DEEP2 Redshift Survey.  They define this line as $^0(U-B)=-0.032(^0\!M_B - 5\log(h/0.7)) - 0.585$, where we have included small corrections to reflect the AB--to--Vega magnitude conversions from \texttt{kcorrect} that have been employed in this study.  We find that our results place the MW redder than this line by $0.067\pm0.071$ mag.

It is interesting to note that in the color--magnitude plane shown in Figure \ref{fig:MW_analogs_cmd}, none of the MW analogs appear in the peak of the blue--cloud region where prototypical blue, star--forming spirals reside, which would hint that our Galaxy, too, very likely does not fit that mold.  This is contrary, however, to what one finds in the SFR--\mt\ plane shown in Figure \ref{fig:MW_analogs_M*_SFR}; the vast majority of the MWAS lie in the blue cloud or just below.  Based on its color, if seen from outside, the MW would likely be defined as a member of the green valley.  In $^0(g-r)$, in fact, it is likely very close to the minimum--density region of color space.  However, based on its \mt\ and \sfr, it appears to fall just off the blue cloud, if it is not actually a member of it.  It thus provides a cautionary example: objects may fall in the green--valley region of parameter space for a variety of reasons, especially when only optical (and not UV) color is considered.

\subsection{Comparisons to Earlier Mass--to--light Ratio Measurements} \label{sec:comp_M2L}
In Table \ref{table:M2L} we have presented new estimates of the global stellar mass--to--light ratio, $\Upsilon^\star$, of the MW in SDSS $ugriz$ passbands in the $z$=0 and 0.1 rest frames, as well as these results transformed to Johnson--Cousins $UBVRI$ passbands.  The most relevant study we can compare these to is \citet{Flynn}, which presented direct estimates of $\Upsilon^\star$ for the local Galactic disk by accounting for the mass and luminosity budget in the ``solar cylinder'' (i.e., the column of stellar material at $R_0$).  This work primarily relied on fitting their \textit{Tuorla} Galactic model to data taken in the Hipparcos and Tycho surveys (reaching out to $\sim$200 pc), which was shown to match well with the \textit{Heidelberg} model--independent analysis of the much more shallow ($<$25--50 pc) Catalogue of Nearby Stars. They found $\Upsilon^\star_V=1.5\pm0.2$ $\massunits/L_\odot$, and then used color conversion derived from Hipparcos/Tycho data to obtain $\Upsilon^\star_B=1.4\pm0.2$ $\massunits/L_\odot$ and $\Upsilon^\star_I=1.2\pm0.2$ $\massunits/L_\odot$.  We note that if we were to update these to reflect the solar absolute magnitudes and colors we have employed herein ($\sim$0.03 mag differences), they would increase by $\sim$3\%, well below the 1$\sigma$ uncertainties.  Regardless, we find that these values compare well with our global MW results of $\Upsilon^\star_B=1.89^{+0.78}_{-0.65}$ $\massunits/L_\odot$, $\Upsilon^\star_V=1.86^{+0.69}_{-0.58}$ $\massunits/L_\odot$, and $\Upsilon^\star_I=1.29^{+0.43}_{-0.37}$ $\massunits/L_\odot$, which are larger than but consistent with the corresponding \citeauthor{Flynn} estimates at the $\sim$ 0.7$\sigma$, 0.6$\sigma$, and 0.2$\sigma$ levels, respectively.

One should keep in mind that, whereas the \citeauthor{Flynn} estimates describe the disk itself, our results represent the global (disk+bulge) values and hence are expected to be larger to some extent, especially in the $B$-- and $V$--bands, as they include the contribution from older stars in the Galactic nucleus.  We can illustrate this further, and hence make a more apples--to--apples comparison, by making the following back--of--the--envelope calculation.  First, for the subset of our volume--limited sample that has $f_{\text{deV}}>0.95$, which constitutes $\sim$33,000 highly bulge--dominated or elliptical galaxies, we find a distribution of $\Upsilon^\star_B$ values that is well approximated as a Gaussian described by $4.1\pm0.9$ $\massunits/L_\odot$ (after multiplying by a factor of 1.5 to convert from Kroupa to Salpeter IMF; cf. \citealp{Fukugita98}).  Second, in LN15 we have determined the bulge--to--total ratio of stellar mass in the MW to be $B/T=0.15\pm0.02$.  By combining our $\Upsilon^\star_B$ estimate for spheroidal components with the \citeauthor{Flynn} $\Upsilon^\star_B$ estimate for the Galactic disk, using the LN15 estimate of $B/T$ to calculate a mass--weighted average for both components, we find a global mass--to--light ratio of $\Upsilon^\star_V=1.81\pm0.19$ $\massunits/L_\odot$.  This is in excellent agreement with our result and is consistent with it at the $\sim$0.1$\sigma$ level.  Doing the analogous calculations in the $V$-- and $I$--bands, the remaining \citeauthor{Flynn} disk values correspond to global values of $\Upsilon^\star_V=1.80\pm0.20$ $\massunits/L_\odot$ and $\Upsilon^\star_I=1.39\pm0.18$ $\massunits/L_\odot$, which are again in excellent agreement with our results, consistent with them at the 0.1$\sigma$ and 0.2$\sigma$ levels, respectively.

\subsection{Conclusions and Future Studies} \label{sec:conclusions}
Overall, since the vast majority of spiral galaxies populate the blue cloud (\citealp{Strateva01}; \citealp{Blanton03}; S14), our results imply that the MW ranks amongst the most luminous, yet reddest of spirals in the local universe.  Based on a variety of empirical definitions in the literature, our results show that it is likely that the MW would be classified as a green--valley galaxy if viewed from the outside, generally taken to indicate that it would be in a transitional evolutionary stage.  Again, this is contrary to what we find in the SFR--\mt\ plane shown in Figure \ref{fig:MW_analogs_M*_SFR}, where we find that the MW lies very near, if not on, the main sequence of star--forming galaxies.  Apparently, even when the impact of dust effects is accounted for, the green valley can be misleading when using it to generally characterize the galaxies it contains (cf. S14).  It is safe to say that our Galaxy's SFR is in a state of decline; the MW produces only $\sim$1.65 solar masses of new stars per year, even though it is amongst the brightest and most massive of late types.  Our findings support the emerging consensus view of the MW; one in which it is not the prototypical, blue spiral it was commonly thought to be just a decade ago, but is instead similar to the passive, red spiral population investigated in \citet{Cortese}.  In fact, based on the demographics of late types presented by S14, if our Galaxy truly lies in the green valley then its photometric properties would be representative of only $\sim$19\% of the spiral galaxies in the nearby universe.  It is beyond the scope of this paper to discuss what evolutionary histories may produce an optically red, yet still star--forming MW (or equivalently late types that appear in the green valley), but we refer the reader to \citet{Hammer}, \citet{Yin}, M11, \citet{Mendez11}, \citet{Jin}, and S14 for insightful discussions.

In following papers we will incorporate other well--studied MW parameters into our technique of studying the MW via its analogs, including morphological type, bulge--to--total ratio, and disk scale length.  In doing so, we should be able to constrain other photometric properties that cannot be directly measured (e.g., the central surface brightness and global S\'ersic index), but are commonly measured for other galaxies.  Additionally, we will integrate UV--wavelength data from \textit{GALEX} to more accurately assess whether the MW belongs in the green valley \citep[cf.][]{Wyder}, as well as utilize \textit{WISE} data to investigate its near--IR properties.  The same multi--wavelength estimates for our nearest MW--like neighbor, M31, the Andromeda Galaxy, whose proximity and thus brightness can cause saturation effects in survey data, would also likely improve from the analog analysis method.  Lastly, the sample of MW analogs we have obtained for this work could also be used to explore other properties with new observations; e.g., an ancillary target program is now underway to provide integral field observations for a subsample of the MWAS studied herein as a part of the SDSS--IV MaNGA survey \citep{Bundy15}.  In a soon--to--follow paper, we will use the results of this paper and those from LN15 to explore the position of the MW in a variety of scaling relations for disk galaxies.

\section*{Acknowledgements}
We are grateful to Simon Mutch, Matt Bershady, Harry Ferguson, Sandra Faber, Leo Blitz, Renbin Yan, Andrew Hearin, Andrew Zentner, Samir Salim, Rachel Somerville, and Risa Wechsler for helpful discussions, and to Genevieve Graves for providing the SDSS color division line used in this paper.  We thank the anonymous referee for suggesting a number of improvements that have been made to this study.  T.C.L. and J.A.N. are supported by the National Science Foundation (NSF) through grant NSF AST 08--06732.  Funding for SDSS--III has been provided by the Alfred P.  Sloan Foundation, the Participating Institutions, the National Science Foundation, and the U.S.  Department of Energy Office of Science.  The SDSS--III web site is \url{http://www.sdss3.org/}.

SDSS--III is managed by the Astrophysical Research Consortium for the Participating Institutions of the SDSS--III Collaboration including the University of Arizona, the Brazilian Participation Group, Brookhaven National Laboratory, University of Cambridge, Carnegie Mellon University, University of Florida, the French Participation Group, the German Participation Group, Harvard University, the Instituto de Astrofisica de Canarias, the Michigan State/Notre Dame/JINA Participation Group, Johns Hopkins University, Lawrence Berkeley National Laboratory, Max Planck Institute for Astrophysics, Max Planck Institute for Extraterrestrial Physics, New Mexico State University, New York University, Ohio State University, Pennsylvania State University, University of Portsmouth, Princeton University, the Spanish Participation Group, University of Tokyo, University of Utah, Vanderbilt University, University of Virginia, University of Washington, and Yale University.  

\bibliographystyle{apj} \footnotesize
\bibliography{mwcolor_v2}

\begin{thebibliography}{}
\expandafter\ifx\csname natexlab\endcsname\relax\def\natexlab#1{#1}\fi

\bibitem[{{Aihara} {et~al.}(2011){Aihara}, {Allende Prieto}, {An}, {Anderson},
  {Aubourg}, {Balbinot}, {Beers}, {Berlind}, {Bickerton}, {Bizyaev}, {Blanton},
  {Bochanski}, {Bolton}, {Bovy}, {Brandt}, {Brinkmann}, {Brown}, {Brownstein},
  {Busca}, {Campbell}, {Carr}, {Chen}, {Chiappini}, {Comparat}, {Connolly},
  {Cortes}, {Croft}, {Cuesta}, {da Costa}, {Davenport}, {Dawson}, {Dhital},
  {Ealet}, {Ebelke}, {Edmondson}, {Eisenstein}, {Escoffier}, {Esposito},
  {Evans}, {Fan}, {Femen{\'{\i}}a Castell{\'a}}, {Font-Ribera}, {Frinchaboy},
  {Ge}, {Gillespie}, {Gilmore}, {Gonz{\'a}lez Hern{\'a}ndez}, {Gott}, {Gould},
  {Grebel}, {Gunn}, {Hamilton}, {Harding}, {Harris}, {Hawley}, {Hearty}, {Ho},
  {Hogg}, {Holtzman}, {Honscheid}, {Inada}, {Ivans}, {Jiang}, {Johnson},
  {Jordan}, {Jordan}, {Kazin}, {Kirkby}, {Klaene}, {Knapp}, {Kneib},
  {Kochanek}, {Koesterke}, {Kollmeier}, {Kron}, {Lampeitl}, {Lang}, {Le Goff},
  {Lee}, {Lin}, {Long}, {Loomis}, {Lucatello}, {Lundgren}, {Lupton}, {Ma},
  {MacDonald}, {Mahadevan}, {Maia}, {Makler}, {Malanushenko}, {Malanushenko},
  {Mandelbaum}, {Maraston}, {Margala}, {Masters}, {McBride}, {McGehee},
  {McGreer}, {M{\'e}nard}, {Miralda-Escud{\'e}}, {Morrison}, {Mullally},
  {Muna}, {Munn}, {Murayama}, {Myers}, {Naugle}, {Neto}, {Nguyen}, {Nichol},
  {O'Connell}, {Ogando}, {Olmstead}, {Oravetz}, {Padmanabhan},
  {Palanque-Delabrouille}, {Pan}, {Pandey}, {P{\^a}ris}, {Percival},
  {Petitjean}, {Pfaffenberger}, {Pforr}, {Phleps}, {Pichon}, {Pieri}, {Prada},
  {Price-Whelan}, {Raddick}, {Ramos}, {Reyl{\'e}}, {Rich}, {Richards}, {Rix},
  {Robin}, {Rocha-Pinto}, {Rockosi}, {Roe}, {Rollinde}, {Ross}, {Ross},
  {Rossetto}, {S{\'a}nchez}, {Sayres}, {Schlegel}, {Schlesinger}, {Schmidt},
  {Schneider}, {Sheldon}, {Shu}, {Simmerer}, {Simmons}, {Sivarani}, {Snedden},
  {Sobeck}, {Steinmetz}, {Strauss}, {Szalay}, {Tanaka}, {Thakar}, {Thomas},
  {Tinker}, {Tofflemire}, {Tojeiro}, {Tremonti}, {Vandenberg}, {Vargas
  Maga{\~n}a}, {Verde}, {Vogt}, {Wake}, {Wang}, {Weaver}, {Weinberg}, {White},
  {White}, {Yanny}, {Yasuda}, {Yeche}, \& {Zehavi}}]{DR8}
{Aihara}, H., {Allende Prieto}, C., {An}, D., {et~al.} 2011, \apjs, 193, 29

\bibitem[{{Bahcall}(1986)}]{Bahcall86}
{Bahcall}, J.~N. 1986, \araa, 24, 577

\bibitem[{{Bahcall} \& {Soneira}(1980)}]{BahcallSoneira}
{Bahcall}, J.~N., \& {Soneira}, R.~M. 1980, \apjs, 44, 73 (B\&S)

\bibitem[{{Baldry} {et~al.}(2006){Baldry}, {Balogh}, {Bower}, {Glazebrook},
  {Nichol}, {Bamford}, \& {Budavari}}]{Baldry}
{Baldry}, I.~K., {Balogh}, M.~L., {Bower}, R.~G., {et~al.} 2006, \mnras, 373,
  469

\bibitem[{{Bell} \& {de Jong}(2001)}]{BelldeJong}
{Bell}, E.~F., \& {de Jong}, R.~S. 2001, \apj, 550, 212

\bibitem[{{Blanton} \& {Roweis}(2007)}]{kcorrect}
{Blanton}, M.~R., \& {Roweis}, S. 2007, \aj, 133, 734

\bibitem[{{Blanton} {et~al.}(2003{\natexlab{a}}){Blanton}, {Hogg}, {Bahcall},
  {Baldry}, {Brinkmann}, {Csabai}, {Eisenstein}, {Fukugita}, {Gunn},
  {Ivezi{\'c}}, {Lamb}, {Lupton}, {Loveday}, {Munn}, {Nichol}, {Okamura},
  {Schlegel}, {Shimasaku}, {Strauss}, {Vogeley}, \& {Weinberg}}]{Blanton03}
{Blanton}, M.~R., {Hogg}, D.~W., {Bahcall}, N.~A., {et~al.} 2003{\natexlab{a}},
  \apj, 594, 186

\bibitem[{{Blanton} {et~al.}(2003{\natexlab{b}}){Blanton}, {Hogg}, {Bahcall},
  {Brinkmann}, {Britton}, {Connolly}, {Csabai}, {Fukugita}, {Loveday},
  {Meiksin}, {Munn}, {Nichol}, {Okamura}, {Quinn}, {Schneider}, {Shimasaku},
  {Strauss}, {Tegmark}, {Vogeley}, \& {Weinberg}}]{BlantonL*}
---. 2003{\natexlab{b}}, \apj, 592, 819

\bibitem[{{Bottinelli} {et~al.}(1985){Bottinelli}, {Gouguenheim}, {Paturel}, \&
  {de Vaucouleurs}}]{sosies1}
{Bottinelli}, L., {Gouguenheim}, L., {Paturel}, G., \& {de Vaucouleurs}, G.
  1985, \apjs, 59, 293

\bibitem[{{Bovy} \& {Rix}(2013)}]{Bovy2013}
{Bovy}, J., \& {Rix}, H.-W. 2013, \apj, 779, 115

\bibitem[{{Brinchmann} {et~al.}(2004){Brinchmann}, {Charlot}, {White},
  {Tremonti}, {Kauffmann}, {Heckman}, \& {Brinkmann}}]{Brinchmann2004}
{Brinchmann}, J., {Charlot}, S., {White}, S.~D.~M., {et~al.} 2004, \mnras, 351,
  1151

\bibitem[{{Bruzual} \& {Charlot}(2003)}]{BC03}
{Bruzual}, G., \& {Charlot}, S. 2003, \mnras, 344, 1000

\bibitem[{{Bundy} {et~al.}(2015){Bundy}, {Bershady}, {Law}, {Yan}, {Drory},
  {MacDonald}, {Wake}, {Cherinka}, {S{\'a}nchez-Gallego}, {Weijmans}, {Thomas},
  {Tremonti}, {Masters}, {Coccato}, {Diamond-Stanic}, {Arag{\'o}n-Salamanca},
  {Avila-Reese}, {Badenes}, {Falc{\'o}n-Barroso}, {Belfiore}, {Bizyaev},
  {Blanc}, {Bland-Hawthorn}, {Blanton}, {Brownstein}, {Byler}, {Cappellari},
  {Conroy}, {Dutton}, {Emsellem}, {Etherington}, {Frinchaboy}, {Fu}, {Gunn},
  {Harding}, {Johnston}, {Kauffmann}, {Kinemuchi}, {Klaene}, {Knapen},
  {Leauthaud}, {Li}, {Lin}, {Maiolino}, {Malanushenko}, {Malanushenko}, {Mao},
  {Maraston}, {McDermid}, {Merrifield}, {Nichol}, {Oravetz}, {Pan}, {Parejko},
  {Sanchez}, {Schlegel}, {Simmons}, {Steele}, {Steinmetz}, {Thanjavur},
  {Thompson}, {Tinker}, {van den Bosch}, {Westfall}, {Wilkinson}, {Wright},
  {Xiao}, \& {Zhang}}]{Bundy15}
{Bundy}, K., {Bershady}, M.~A., {Law}, D.~R., {et~al.} 2015, \apj, 798, 7

\bibitem[{{Cardelli} {et~al.}(1989){Cardelli}, {Clayton}, \&
  {Mathis}}]{Cardelli}
{Cardelli}, J.~A., {Clayton}, G.~C., \& {Mathis}, J.~S. 1989, \apj, 345, 245

\bibitem[{{Charlot} \& {Longhetti}(2001)}]{CL01}
{Charlot}, S., \& {Longhetti}, M. 2001, \mnras, 323, 887

\bibitem[{{Chatzopoulos} {et~al.}(2015){Chatzopoulos}, {Fritz}, {Gerhard},
  {Gillessen}, {Wegg}, {Genzel}, \& {Pfuhl}}]{Chatzo}
{Chatzopoulos}, S., {Fritz}, T.~K., {Gerhard}, O., {et~al.} 2015, \mnras, 447,
  952

\bibitem[{{Chomiuk} \& {Povich}(2011)}]{Chomiuk}
{Chomiuk}, L., \& {Povich}, M.~S. 2011, \aj, 142, 197

\bibitem[{{Cook} {et~al.}(2014){Cook}, {Dale}, {Johnson}, {Van Zee}, {Lee},
  {Kennicutt}, {Calzetti}, {Staudaher}, \& {Engelbracht}}]{Cook}
{Cook}, D.~O., {Dale}, D.~A., {Johnson}, B.~D., {et~al.} 2014, \mnras, 445, 890

\bibitem[{{Cortese}(2012)}]{Cortese}
{Cortese}, L. 2012, \aap, 543, A132

\bibitem[{{Dawson} {et~al.}(2013){Dawson}, {Schlegel}, {Ahn}, {Anderson},
  {Aubourg}, {Bailey}, {Barkhouser}, {Bautista}, {Beifiori}, {Berlind},
  {Bhardwaj}, {Bizyaev}, {Blake}, {Blanton}, {Blomqvist}, {Bolton}, {Borde},
  {Bovy}, {Brandt}, {Brewington}, {Brinkmann}, {Brown}, {Brownstein}, {Bundy},
  {Busca}, {Carithers}, {Carnero}, {Carr}, {Chen}, {Comparat}, {Connolly},
  {Cope}, {Croft}, {Cuesta}, {da Costa}, {Davenport}, {Delubac}, {de Putter},
  {Dhital}, {Ealet}, {Ebelke}, {Eisenstein}, {Escoffier}, {Fan}, {Filiz Ak},
  {Finley}, {Font-Ribera}, {G{\'e}nova-Santos}, {Gunn}, {Guo}, {Haggard},
  {Hall}, {Hamilton}, {Harris}, {Harris}, {Ho}, {Hogg}, {Holder}, {Honscheid},
  {Huehnerhoff}, {Jordan}, {Jordan}, {Kauffmann}, {Kazin}, {Kirkby}, {Klaene},
  {Kneib}, {Le Goff}, {Lee}, {Long}, {Loomis}, {Lundgren}, {Lupton}, {Maia},
  {Makler}, {Malanushenko}, {Malanushenko}, {Mandelbaum}, {Manera}, {Maraston},
  {Margala}, {Masters}, {McBride}, {McDonald}, {McGreer}, {McMahon}, {Mena},
  {Miralda-Escud{\'e}}, {Montero-Dorta}, {Montesano}, {Muna}, {Myers},
  {Naugle}, {Nichol}, {Noterdaeme}, {Nuza}, {Olmstead}, {Oravetz}, {Oravetz},
  {Owen}, {Padmanabhan}, {Palanque-Delabrouille}, {Pan}, {Parejko},
  {P{\^a}ris}, {Percival}, {P{\'e}rez-Fournon}, {P{\'e}rez-R{\`a}fols},
  {Petitjean}, {Pfaffenberger}, {Pforr}, {Pieri}, {Prada}, {Price-Whelan},
  {Raddick}, {Rebolo}, {Rich}, {Richards}, {Rockosi}, {Roe}, {Ross}, {Ross},
  {Rossi}, {Rubi{\~n}o-Martin}, {Samushia}, {S{\'a}nchez}, {Sayres}, {Schmidt},
  {Schneider}, {Sc{\'o}ccola}, {Seo}, {Shelden}, {Sheldon}, {Shen}, {Shu},
  {Slosar}, {Smee}, {Snedden}, {Stauffer}, {Steele}, {Strauss}, {Streblyanska},
  {Suzuki}, {Swanson}, {Tal}, {Tanaka}, {Thomas}, {Tinker}, {Tojeiro},
  {Tremonti}, {Vargas Maga{\~n}a}, {Verde}, {Viel}, {Wake}, {Watson}, {Weaver},
  {Weinberg}, {Weiner}, {West}, {White}, {Wood-Vasey}, {Yeche}, {Zehavi},
  {Zhao}, \& {Zheng}}]{Dawson}
{Dawson}, K.~S., {Schlegel}, D.~J., {Ahn}, C.~P., {et~al.} 2013, \aj, 145, 10

\bibitem[{{de Vaucouleurs}(1970)}]{deV1970}
{de Vaucouleurs}, G. 1970, \apj, 159, 435

\bibitem[{{de Vaucouleurs}(1977)}]{dV1977}
{de Vaucouleurs}, G. 1977, in Evolution of Galaxies and Stellar Populations,
  ed. B.~M. {Tinsley} \& R.~B.~G. {Larson}, D.~Campbell (New Haven, CT: Yale
  Univ. Observatory), 43 (dV77)

\bibitem[{{de Vaucouleurs}(1983)}]{deV1983}
---. 1983, \apj, 268, 451 (dV83)

\bibitem[{{de Vaucouleurs} \& {Corwin}(1986)}]{sosies2}
{de Vaucouleurs}, G., \& {Corwin}, Jr., H.~G. 1986, \apj, 308, 487

\bibitem[{{de Vaucouleurs} {et~al.}(1976){de Vaucouleurs}, {de Vaucouleurs}, \&
  {Corwin}}]{RC2}
{de Vaucouleurs}, G., {de Vaucouleurs}, A., \& {Corwin}, Jr., H.~G. 1976,
  {Second Reference Catalogue of Bright Galaxies} (Austin, TX: Univ. Texas
  Press)

\bibitem[{{de Vaucouleurs} {et~al.}(1991){de Vaucouleurs}, {de Vaucouleurs},
  {Corwin}, {Buta}, {Paturel}, \& {Fouqu{\'e}}}]{RC3}
{de Vaucouleurs}, G., {de Vaucouleurs}, A., {Corwin}, Jr., H.~G., {et~al.}
  1991, {Third Reference Catalogue of Bright Galaxies} (Berlin: Springer)

\bibitem[{{de Vaucouleurs} \& {Pence}(1978)}]{deVPence}
{de Vaucouleurs}, G., \& {Pence}, W.~D. 1978, \aj, 83, 1163 (dV\&P)

\bibitem[{Efron(1979)}]{efron}
Efron, B. 1979, AnSta, 7, 1

\bibitem[{{Faber} {et~al.}(2007){Faber}, {Willmer}, {Wolf}, {Koo}, {Weiner},
  {Newman}, {Im}, {Coil}, {Conroy}, {Cooper}, {Davis}, {Finkbeiner}, {Gerke},
  {Gebhardt}, {Groth}, {Guhathakurta}, {Harker}, {Kaiser}, {Kassin},
  {Kleinheinrich}, {Konidaris}, {Kron}, {Lin}, {Luppino}, {Madgwick},
  {Meisenheimer}, {Noeske}, {Phillips}, {Sarajedini}, {Schiavon}, {Simard},
  {Szalay}, {Vogt}, \& {Yan}}]{Faber07}
{Faber}, S.~M., {Willmer}, C.~N.~A., {Wolf}, C., {et~al.} 2007, \apj, 665, 265

\bibitem[{{Fang} {et~al.}(2012){Fang}, {Faber}, {Salim}, {Graves}, \&
  {Rich}}]{Fang}
{Fang}, J.~J., {Faber}, S.~M., {Salim}, S., {Graves}, G.~J., \& {Rich}, R.~M.
  2012, \apj, 761, 23

\bibitem[{{Fern{\'a}ndez Lorenzo} {et~al.}(2012){Fern{\'a}ndez Lorenzo},
  {Sulentic}, {Verdes-Montenegro}, {Ruiz}, {Sabater}, \&
  {S{\'a}nchez}}]{Lorenzo2012}
{Fern{\'a}ndez Lorenzo}, M., {Sulentic}, J., {Verdes-Montenegro}, L., {et~al.}
  2012, \aap, 540, A47

\bibitem[{{Flynn} {et~al.}(2006){Flynn}, {Holmberg}, {Portinari}, {Fuchs}, \&
  {Jahrei{\ss}}}]{Flynn}
{Flynn}, C., {Holmberg}, J., {Portinari}, L., {Fuchs}, B., \& {Jahrei{\ss}}, H.
  2006, \mnras, 372, 1149

\bibitem[{{Fukugita} {et~al.}(1998){Fukugita}, {Hogan}, \&
  {Peebles}}]{Fukugita98}
{Fukugita}, M., {Hogan}, C.~J., \& {Peebles}, P.~J.~E. 1998, \apj, 503, 518

\bibitem[{{Fukugita} {et~al.}(1995){Fukugita}, {Shimasaku}, \&
  {Ichikawa}}]{Fukugita95}
{Fukugita}, M., {Shimasaku}, K., \& {Ichikawa}, T. 1995, \pasp, 107, 945

\bibitem[{{Gallazzi} {et~al.}(2005){Gallazzi}, {Charlot}, {Brinchmann},
  {White}, \& {Tremonti}}]{Gallazzi05}
{Gallazzi}, A., {Charlot}, S., {Brinchmann}, J., {White}, S.~D.~M., \&
  {Tremonti}, C.~A. 2005, \mnras, 362, 41

\bibitem[{{Gillessen} {et~al.}(2009){Gillessen}, {Eisenhauer}, {Trippe},
  {Alexander}, {Genzel}, {Martins}, \& {Ott}}]{Gillessen}
{Gillessen}, S., {Eisenhauer}, F., {Trippe}, S., {et~al.} 2009, \apj, 692, 1075

\bibitem[{{Gon{\c c}alves} {et~al.}(2012){Gon{\c c}alves}, {Martin},
  {Men{\'e}ndez-Delmestre}, {Wyder}, \& {Koekemoer}}]{Gon}
{Gon{\c c}alves}, T.~S., {Martin}, D.~C., {Men{\'e}ndez-Delmestre}, K.,
  {Wyder}, T.~K., \& {Koekemoer}, A. 2012, \apj, 759, 67

\bibitem[{{Graves}(2012)}]{Graves}
{Graves}, G. 2012, private communication

\bibitem[{{Hammer} {et~al.}(2007){Hammer}, {Puech}, {Chemin}, {Flores}, \&
  {Lehnert}}]{Hammer}
{Hammer}, F., {Puech}, M., {Chemin}, L., {Flores}, H., \& {Lehnert}, M.~D.
  2007, \apj, 662, 322

\bibitem[{{Herschel}(1785)}]{Herschel}
{Herschel}, S.~W. 1785, Phil. Trans. R. Soc. London, 75, 213

\bibitem[{Hodges \& Lehmann(1963)}]{HLMean}
Hodges, Jr., J.~L., \& Lehmann, E.~L. 1963, Ann. Math. Statist., 34, 598

\bibitem[{{Hogg} {et~al.}(2002){Hogg}, {Baldry}, {Blanton}, \&
  {Eisenstein}}]{HoggKcorrection}
{Hogg}, D.~W., {Baldry}, I.~K., {Blanton}, M.~R., \& {Eisenstein}, D.~J. 2002,
  ArXiv Astrophysics e-prints, astro-ph/0210394

\bibitem[{{Ilbert} {et~al.}(2010){Ilbert}, {Salvato}, {Le Floc'h}, {Aussel},
  {Capak}, {McCracken}, {Mobasher}, {Kartaltepe}, {Scoville}, {Sanders},
  {Arnouts}, {Bundy}, {Cassata}, {Kneib}, {Koekemoer}, {Le F{\`e}vre}, {Lilly},
  {Surace}, {Taniguchi}, {Tasca}, {Thompson}, {Tresse}, {Zamojski}, {Zamorani},
  \& {Zucca}}]{Ilbert10}
{Ilbert}, O., {Salvato}, M., {Le Floc'h}, E., {et~al.} 2010, \apj, 709, 644

\bibitem[{{Jin} {et~al.}(2014){Jin}, {Gu}, {Huang}, {Shi}, \& {Feng}}]{Jin}
{Jin}, S.-W., {Gu}, Q., {Huang}, S., {Shi}, Y., \& {Feng}, L.-L. 2014, \apj,
  787, 63

\bibitem[{{Kauffmann} {et~al.}(2003){Kauffmann}, {Heckman}, {White}, {Charlot},
  {Tremonti}, {Brinchmann}, {Bruzual}, {Peng}, {Seibert}, {Bernardi},
  {Blanton}, {Brinkmann}, {Castander}, {Cs{\'a}bai}, {Fukugita}, {Ivezic},
  {Munn}, {Nichol}, {Padmanabhan}, {Thakar}, {Weinberg}, \& {York}}]{Kauffmann}
{Kauffmann}, G., {Heckman}, T.~M., {White}, S.~D.~M., {et~al.} 2003, \mnras,
  341, 33

\bibitem[{{Kennicutt}(1998)}]{Kennicutt98}
{Kennicutt}, Jr., R.~C. 1998, \apj, 498, 541

\bibitem[{{Kreiken}(1950)}]{Kreiken}
{Kreiken}, E.~A. 1950, {Some Remarks on the Surface Brightness of the Stellar
  System} (Indonesia: OSR)

\bibitem[{{Kroupa} \& {Weidner}(2003)}]{Kroupa}
{Kroupa}, P., \& {Weidner}, C. 2003, \apj, 598, 1076

\bibitem[{{Leauthaud} {et~al.}(2012){Leauthaud}, {Tinker}, {Bundy}, {Behroozi},
  {Massey}, {Rhodes}, {George}, {Kneib}, {Benson}, {Wechsler}, {Busha},
  {Capak}, {Cort{\^e}s}, {Ilbert}, {Koekemoer}, {Le F{\`e}vre}, {Lilly},
  {McCracken}, {Salvato}, {Schrabback}, {Scoville}, {Smith}, \&
  {Taylor}}]{Leauthaud12}
{Leauthaud}, A., {Tinker}, J., {Bundy}, K., {et~al.} 2012, \apj, 744, 159

\bibitem[{{Licquia} \& {Newman}(2015)}]{Licquia}
{Licquia}, T.~C., \& {Newman}, J.~A. 2015, \apj, 806, 96 (LN15)

\bibitem[{{Liu} {et~al.}(2011){Liu}, {Gerke}, {Wechsler}, {Behroozi}, \&
  {Busha}}]{Liu}
{Liu}, L., {Gerke}, B.~F., {Wechsler}, R.~H., {Behroozi}, P.~S., \& {Busha},
  M.~T. 2011, \apj, 733, 62

\bibitem[{{Maller} {et~al.}(2009){Maller}, {Berlind}, {Blanton}, \&
  {Hogg}}]{Maller}
{Maller}, A.~H., {Berlind}, A.~A., {Blanton}, M.~R., \& {Hogg}, D.~W. 2009,
  \apj, 691, 394

\bibitem[{{Mendez} {et~al.}(2011){Mendez}, {Coil}, {Lotz}, {Salim},
  {Moustakas}, \& {Simard}}]{Mendez11}
{Mendez}, A.~J., {Coil}, A.~L., {Lotz}, J., {et~al.} 2011, \apj, 736, 110

\bibitem[{{Montero-Dorta} \& {Prada}(2009)}]{Dorta}
{Montero-Dorta}, A.~D., \& {Prada}, F. 2009, \mnras, 399, 1106

\bibitem[{{Mutch} {et~al.}(2011){Mutch}, {Croton}, \& {Poole}}]{Mutch}
{Mutch}, S.~J., {Croton}, D.~J., \& {Poole}, G.~B. 2011, \apj, 736, 84 (M11)

\bibitem[{{Reach} {et~al.}(1996){Reach}, {Abergel}, {Boulanger}, {Desert},
  {Perault}, {Bernard}, {Blommaert}, {Cesarsky}, {Cesarsky}, {Metcalfe},
  {Puget}, {Sibille}, \& {Vigroux}}]{Reach}
{Reach}, W.~T., {Abergel}, A., {Boulanger}, F., {et~al.} 1996, \aap, 315, L381

\bibitem[{{Salim} {et~al.}(2007){Salim}, {Rich}, {Charlot}, {Brinchmann},
  {Johnson}, {Schiminovich}, {Seibert}, {Mallery}, {Heckman}, {Forster},
  {Friedman}, {Martin}, {Morrissey}, {Neff}, {Small}, {Wyder}, {Bianchi},
  {Donas}, {Lee}, {Madore}, {Milliard}, {Szalay}, {Welsh}, \& {Yi}}]{Salim}
{Salim}, S., {Rich}, R.~M., {Charlot}, S., {et~al.} 2007, \apjs, 173, 267

\bibitem[{{Salim} {et~al.}(2009){Salim}, {Dickinson}, {Michael Rich},
  {Charlot}, {Lee}, {Schiminovich}, {P{\'e}rez-Gonz{\'a}lez}, {Ashby},
  {Papovich}, {Faber}, {Ivison}, {Frayer}, {Walton}, {Weiner}, {Chary},
  {Bundy}, {Noeske}, \& {Koekemoer}}]{Salim09}
{Salim}, S., {Dickinson}, M., {Michael Rich}, R., {et~al.} 2009, \apj, 700, 161

\bibitem[{{Schawinski} {et~al.}(2014){Schawinski}, {Urry}, {Simmons},
  {Fortson}, {Kaviraj}, {Keel}, {Lintott}, {Masters}, {Nichol}, {Sarzi},
  {Skibba}, {Treister}, {Willett}, {Wong}, \& {Yi}}]{Schawinski}
{Schawinski}, K., {Urry}, C.~M., {Simmons}, B.~D., {et~al.} 2014, \mnras, 440,
  889 (S14)

\bibitem[{{Schlegel} {et~al.}(1998){Schlegel}, {Finkbeiner}, \& {Davis}}]{SFD}
{Schlegel}, D.~J., {Finkbeiner}, D.~P., \& {Davis}, M. 1998, \apj, 500, 525

\bibitem[{{Schmidt-Kaler} \& {Schlosser}(1973)}]{SchmidtKaler}
{Schmidt-Kaler}, T., \& {Schlosser}, W. 1973, \aap, 29, 409

\bibitem[{{Strateva} {et~al.}(2001){Strateva}, {Ivezi{\'c}}, {Knapp},
  {Narayanan}, {Strauss}, {Gunn}, {Lupton}, {Schlegel}, {Bahcall}, {Brinkmann},
  {Brunner}, {Budav{\'a}ri}, {Csabai}, {Castander}, {Doi}, {Fukugita}, {Gy{\H
  o}ry}, {Hamabe}, {Hennessy}, {Ichikawa}, {Kunszt}, {Lamb}, {McKay},
  {Okamura}, {Racusin}, {Sekiguchi}, {Schneider}, {Shimasaku}, \&
  {York}}]{Strateva01}
{Strateva}, I., {Ivezi{\'c}}, {\v Z}., {Knapp}, G.~R., {et~al.} 2001, \aj, 122,
  1861

\bibitem[{{Strauss} {et~al.}(2002){Strauss}, {Weinberg}, {Lupton}, {Narayanan},
  {Annis}, {Bernardi}, {Blanton}, {Burles}, {Connolly}, {Dalcanton}, {Doi},
  {Eisenstein}, {Frieman}, {Fukugita}, {Gunn}, {Ivezi{\'c}}, {Kent}, {Kim},
  {Knapp}, {Kron}, {Munn}, {Newberg}, {Nichol}, {Okamura}, {Quinn}, {Richmond},
  {Schlegel}, {Shimasaku}, {SubbaRao}, {Szalay}, {Vanden Berk}, {Vogeley},
  {Yanny}, {Yasuda}, {York}, \& {Zehavi}}]{Strauss}
{Strauss}, M.~A., {Weinberg}, D.~H., {Lupton}, R.~H., {et~al.} 2002, \aj, 124,
  1810

\bibitem[{{Tago} {et~al.}(2010){Tago}, {Saar}, {Tempel}, {Einasto}, {Einasto},
  {Nurmi}, \& {Hein{\"a}m{\"a}ki}}]{Tago10}
{Tago}, E., {Saar}, E., {Tempel}, E., {et~al.} 2010, \aap, 514, A102

\bibitem[{{Taylor} {et~al.}(2015){Taylor}, {Hopkins}, {Baldry},
  {Bland-Hawthorn}, {Brown}, {Colless}, {Driver}, {Norberg}, {Robotham},
  {Alpaslan}, {Brough}, {Cluver}, {Gunawardhana}, {Kelvin}, {Liske},
  {Conselice}, {Croom}, {Foster}, {Jarrett}, {Lara-Lopez}, \&
  {Loveday}}]{Taylor14}
{Taylor}, E.~N., {Hopkins}, A.~M., {Baldry}, I.~K., {et~al.} 2015, \mnras, 446,
  2144

\bibitem[{{Tempel} {et~al.}(2014){Tempel}, {Tamm}, {Gramann}, {Tuvikene},
  {Liivam{\"a}gi}, {Suhhonenko}, {Kipper}, {Einasto}, \& {Saar}}]{Tempel14}
{Tempel}, E., {Tamm}, A., {Gramann}, M., {et~al.} 2014, \aap, 566, A1

\bibitem[{{Unterborn} \& {Ryden}(2008)}]{Inclination}
{Unterborn}, C.~T., \& {Ryden}, B.~S. 2008, \apj, 687, 976

\bibitem[{{van den Bergh}(2000)}]{vandenBergh}
{van den Bergh}, S. 2000, \pasp, 112, 529

\bibitem[{{van der Kruit}(1986)}]{vanderKruit}
{van der Kruit}, P.~C. 1986, \aap, 157, 230 (vdK86)

\bibitem[{Wilk \& Gnanadesikan(1968)}]{Wilk68}
Wilk, M.~B., \& Gnanadesikan, R. 1968, Biometrika, 55, 1

\bibitem[{{Willmer} {et~al.}(2006){Willmer}, {Faber}, {Koo}, {Weiner},
  {Newman}, {Coil}, {Connolly}, {Conroy}, {Cooper}, {Davis}, {Finkbeiner},
  {Gerke}, {Guhathakurta}, {Harker}, {Kaiser}, {Kassin}, {Konidaris}, {Lin},
  {Luppino}, {Madgwick}, {Noeske}, {Phillips}, \& {Yan}}]{Willmer06}
{Willmer}, C.~N.~A., {Faber}, S.~M., {Koo}, D.~C., {et~al.} 2006, \apj, 647,
  853

\bibitem[{{Wong} {et~al.}(2012){Wong}, {Schawinski}, {Kaviraj}, {Masters},
  {Nichol}, {Lintott}, {Keel}, {Darg}, {Bamford}, {Andreescu}, {Murray},
  {Raddick}, {Szalay}, {Thomas}, \& {Vandenberg}}]{Wong12}
{Wong}, O.~I., {Schawinski}, K., {Kaviraj}, S., {et~al.} 2012, \mnras, 420,
  1684

\bibitem[{{Wyder} {et~al.}(2007){Wyder}, {Martin}, {Schiminovich}, {Seibert},
  {Budav{\'a}ri}, {Treyer}, {Barlow}, {Forster}, {Friedman}, {Morrissey},
  {Neff}, {Small}, {Bianchi}, {Donas}, {Heckman}, {Lee}, {Madore}, {Milliard},
  {Rich}, {Szalay}, {Welsh}, \& {Yi}}]{Wyder}
{Wyder}, T.~K., {Martin}, D.~C., {Schiminovich}, D., {et~al.} 2007, \apjs, 173,
  293

\bibitem[{{Yin} {et~al.}(2009){Yin}, {Hou}, {Prantzos}, {Boissier}, {Chang},
  {Shen}, \& {Zhang}}]{Yin}
{Yin}, J., {Hou}, J.~L., {Prantzos}, N., {et~al.} 2009, \aap, 505, 497

\bibitem[{{York} {et~al.}(2000){York}, {Adelman}, {Anderson}, {Anderson},
  {Annis}, {Bahcall}, {Bakken}, {Barkhouser}, {Bastian}, {Berman}, {Boroski},
  {Bracker}, {Briegel}, {Briggs}, {Brinkmann}, {Brunner}, {Burles}, {Carey},
  {Carr}, {Castander}, {Chen}, {Colestock}, {Connolly}, {Crocker}, {Csabai},
  {Czarapata}, {Davis}, {Doi}, {Dombeck}, {Eisenstein}, {Ellman}, {Elms},
  {Evans}, {Fan}, {Federwitz}, {Fiscelli}, {Friedman}, {Frieman}, {Fukugita},
  {Gillespie}, {Gunn}, {Gurbani}, {de Haas}, {Haldeman}, {Harris}, {Hayes},
  {Heckman}, {Hennessy}, {Hindsley}, {Holm}, {Holmgren}, {Huang}, {Hull},
  {Husby}, {Ichikawa}, {Ichikawa}, {Ivezi{\'c}}, {Kent}, {Kim}, {Kinney},
  {Klaene}, {Kleinman}, {Kleinman}, {Knapp}, {Korienek}, {Kron}, {Kunszt},
  {Lamb}, {Lee}, {Leger}, {Limmongkol}, {Lindenmeyer}, {Long}, {Loomis},
  {Loveday}, {Lucinio}, {Lupton}, {MacKinnon}, {Mannery}, {Mantsch}, {Margon},
  {McGehee}, {McKay}, {Meiksin}, {Merelli}, {Monet}, {Munn}, {Narayanan},
  {Nash}, {Neilsen}, {Neswold}, {Newberg}, {Nichol}, {Nicinski}, {Nonino},
  {Okada}, {Okamura}, {Ostriker}, {Owen}, {Pauls}, {Peoples}, {Peterson},
  {Petravick}, {Pier}, {Pope}, {Pordes}, {Prosapio}, {Rechenmacher}, {Quinn},
  {Richards}, {Richmond}, {Rivetta}, {Rockosi}, {Ruthmansdorfer}, {Sandford},
  {Schlegel}, {Schneider}, {Sekiguchi}, {Sergey}, {Shimasaku}, {Siegmund},
  {Smee}, {Smith}, {Snedden}, {Stone}, {Stoughton}, {Strauss}, {Stubbs},
  {SubbaRao}, {Szalay}, {Szapudi}, {Szokoly}, {Thakar}, {Tremonti}, {Tucker},
  {Uomoto}, {Vanden Berk}, {Vogeley}, {Waddell}, {Wang}, {Watanabe},
  {Weinberg}, {Yanny}, {Yasuda}, \& {SDSS Collaboration}}]{York2000}
{York}, D.~G., {Adelman}, J., {Anderson}, Jr., J.~E., {et~al.} 2000, \aj, 120,
  1579

\end{thebibliography}

\end{document}